\documentclass[aps,amsmath,preprint,epsf,superscriptaddress,nofootinbib]{revtex4-1}

\usepackage[dvipdfmx]{graphicx}
\usepackage{epstopdf}
\usepackage{dcolumn}
\usepackage{bm}
\usepackage{hyperref}
\usepackage{color}
\usepackage{cancel}
\usepackage{bbm}
 \usepackage{mathtools}
 
 \usepackage{tcolorbox}
\usepackage{mathtools}

\newcommand{\defeq}{=}

\let\oldfootnote\footnote
\renewcommand{\footnote}[1]{%
  \begingroup%
  \linespread{1}%
  \oldfootnote{#1}%
  \endgroup%
}

\newcolumntype{C}{>{\centering\arraybackslash}p{1.cm}}

\newcommand{\tj}[6]{ 
\begin{pmatrix}
   #1 & #2 & #3 \\
   #4 & #5 & #6 
\end{pmatrix}
}

\begin{document}


\def\a{\alpha}
\def\b{\beta}
\def\c{\varepsilon}
\def\d{\delta}
\def\e{\epsilon}
\def\f{\phi}
\def\g{\gamma}
\def\h{\theta}
\def\k{\kappa}
\def\l{\lambda}
\def\m{\mu}
\def\n{\nu}
\def\p{\psi}
\def\q{\partial}
\def\r{\rho}
\def\s{\sigma}
\def\t{\tau}
\def\u{\upsilon}
\def\v{\varphi}
\def\w{\omega}
\def\x{\xi}
\def\y{\eta}
\def\z{\zeta}
\def\D{\Delta}
\def\G{\Gamma}
\def\H{\Theta}
\def\L{\Lambda}
\def\F{\Phi}
\def\P{\Psi}
\def\S{\Sigma}

\def\o{\over}
\def\beq{\begin{eqnarray}}
\def\eeq{\end{eqnarray}}
\newcommand{\gsim}{ \mathop{}_{\textstyle \sim}^{\textstyle >} }
\newcommand{\lsim}{ \mathop{}_{\textstyle \sim}^{\textstyle <} }
\newcommand{\vev}[1]{ \left\langle {#1} \right\rangle }
\newcommand{\bra}[1]{ \langle {#1} | }
\newcommand{\ket}[1]{ | {#1} \rangle }
\newcommand{\EV}{ {\rm eV} }
\newcommand{\KEV}{ {\rm keV} }
\newcommand{\MEV}{ {\rm MeV} }
\newcommand{\GEV}{ {\rm GeV} }
\newcommand{\TEV}{ {\rm TeV} }
\newcommand{\1}{\mbox{1}\hspace{-0.25em}\mbox{l}}
\newcommand{\headline}[1]{\noindent{\bf #1}}
\def\diag{\mathop{\rm diag}\nolimits}
\def\Spin{\mathop{\rm Spin}}
\def\SO{\mathop{\rm SO}}
\def\O{\mathop{\rm O}}
\def\SU{\mathop{\rm SU}}
\def\U{\mathop{\rm U}}
\def\Sp{\mathop{\rm Sp}}
\def\SL{\mathop{\rm SL}}
\def\tr{\mathop{\rm tr}}
\def\mpl{M_{PL}}

\def\IJMP{Int.~J.~Mod.~Phys. }
\def\MPL{Mod.~Phys.~Lett. }
\def\NP{Nucl.~Phys. }
\def\PL{Phys.~Lett. }
\def\PR{Phys.~Rev. }
\def\PRL{Phys.~Rev.~Lett. }
\def\PTP{Prog.~Theor.~Phys. }
\def\ZP{Z.~Phys. }

\def\dd{\mathrm{d}}
\def\ff{\mathrm{f}}
\def\BH{{\rm BH}}
\def\inf{{\rm inf}}
\def\ev{{\rm evap}}
\def\eq{{\rm eq}}
\def\SM{{\rm sm}}
\def\Mpl{M_{\rm Pl}}
\def\GeV{{\rm GeV}}
\newcommand{\Red}[1]{\textcolor{red}{#1}}

\def\RN{\mathbf{R}_N}
\def\RDM{\mathbf{R}_{DM}}
\def\hRN{\hat{\mathbf{R}}_N}
\def\hRDM{\hat{\mathbf{R}}_{DM}}
\def\XN{\mathbf{x}_N}
\def\XDM{\mathbf{x}_{DM}}
\def\hXN{\hat{\mathbf{x}}_N}
\def\hXDM{\hat{\mathbf{x}}_{DM}}
\def\xe{\mathbf{x}}
\def\xei{\mathbf{x}_i}
\def\xej{\mathbf{x}_j}

\newcommand{\prob}{\mathcal{P}}
\newcommand{\nominalMass}{\overline{m}_A}
\newcommand{\physicalMass}{m_A}

\preprint{IPMU17-0100}
\bigskip

\title{Migdal Effect in Dark Matter Direct Detection Experiments}
\author{Masahiro Ibe}
\email[e-mail: ]{ibe@icrr.u-tokyo.ac.jp}
\affiliation{ICRR, The University of Tokyo, Kashiwa, Chiba 277-8582, Japan}
\affiliation{Kavli IPMU (WPI), UTIAS, The University of Tokyo, Kashiwa, Chiba 277-8583, Japan}

\author{Wakutaka Nakano}
\email[e-mail: ]{m156077@icrr.u-tokyo.ac.jp}
\affiliation{ICRR, The University of Tokyo, Kashiwa, Chiba 277-8582, Japan}

\author{Yutaro Shoji}
\email[e-mail: ]{yshoji@icrr.u-tokyo.ac.jp}
\affiliation{ICRR, The University of Tokyo, Kashiwa, Chiba 277-8582, Japan}

\author{Kazumine Suzuki}
\email[e-mail: ]{ksuzuki@icrr.u-tokyo.ac.jp}
\affiliation{ICRR, The University of Tokyo, Kashiwa, Chiba 277-8582, Japan}

\date{\today}

\begin{abstract}
The elastic scattering of an atomic nucleus plays a central role in dark matter direct detection experiments.
In those experiments, it is usually assumed that the atomic electrons around the nucleus of the target material immediately follow the motion of the recoil nucleus.
In reality, however, it takes some time for the electrons to catch up, which results in ionization and excitation of the atoms.
In previous studies, those effects are taken into account by using the so-called Migdal's approach,  
in which the final state ionization/excitation are treated separately from the nuclear recoil.
In this paper, we reformulate the Migdal's approach so that the ``atomic recoil'' cross section
is obtained coherently, where we make transparent the energy-momentum conservation and the probability conservation.
We show that the final state ionization/excitation can enhance the detectability of 
rather light dark matter in the GeV mass range via the {\it nuclear} scattering.
We also discuss the coherent neutrino-nucleus  scattering, where the same effects are expected.
\end{abstract}

\maketitle
\section{Introdcution}
The existence of dark matter is overwhelmingly supported by numerous cosmological and astrophysical 
observations on a wide range of scales.
However, the nature of dark matter has not been revealed for almost a century except for its gravitational interactions.
The identification of the nature of dark matter is one of the most important challenges of modern particle physics (see~\cite{Bertone:2004pz,Murayama:2007ek,Feng:2010gw} for review).

Among various candidates for dark matter, the weakly interacting massive particles (WIMPs) are the most extensively studied category of dark matter.
The WIMPs couple to the standard model particles via interactions similar in strength to the weak nuclear force.
Through the weak interaction, the WIMPs  are thermally produced in the early universe,
and the relic density is set as they freeze out from the thermal bath~\cite{Lee:1977ua}. 
The WIMPs are particularly attractive since the dark matter density does not depend on the details of the initial condition 
of the universe.
The WIMPs are also highly motivated as they are interrelated to physics beyond the standard model
such as supersymmetry (see e.g.~\cite{Jungman:1995df}).

The ambient WIMPs can be directly detected by searching for its scattering with the atomic nuclei~\cite{Goodman:1984dc}.
Given a circular speed at around the Sun of $239 \pm 5$\,km/s~\cite{Nesti:2013uwa}, 
the WIMPs recoil the nuclei elastically with a typical momentum transfer of $q_A \sim 100$\,MeV
for the target nucleus mass of $m_N \sim 100$\,GeV.%
\footnote{For a lighter WIMP than $m_N$, $q_A$ is suppressed by  the reduced mass between the nucleus and the WIMP.
}
The recoil signatures are detected through ionization, scintillation, and the production of heat
in the detectors (see~\cite{Lewin:1995rx,Gaitskell:2004gd,Undagoitia:2015gya} for review).
To date, for example, liquid xenon detectors 
such as LUX~\cite{Akerib:2016vxi}, PandaX-II~\cite{Tan:2016zwf}, and XENON1T~\cite{Aprile:2017iyp} 
have put stringent exclusion limits on the spin-independent WIMP-nucleon recoil cross section.

In those experiments, it is usually assumed that 
the atomic electrons around the recoil nucleus immediately follow the motion of the nucleus.
However, it takes some time for the electrons to catch up,
which causes ionization and excitation of the recoil atom.
The ionization and the excitation result in extra electronic energy injections into the detectors.%
\footnote{The rates of the ionization/excitation are much smaller than ${\cal O}(1)$.
Besides, the same effects are expected in the nuclear recoil by neutron injections.
Accordingly, those effects are almost always taken into account in detector calibration by the neutron sources automatically.}
The importance of such effects on direct detection experiments 
has been pointed out~\cite{Vergados:2004bm,*Moustakidis:2005gx,*Ejiri:2005aj,*Vergados:2013raa,Bernabei:2007jz}.
(See also Refs.\,\cite{Roberts:2015lga,Roberts:2016xfw}, which discuss the ionization effects in the direct detection experiments
for dark matter electron scattering.)

In previous studies, such effects are estimated by using the so-called Migdal's approach~\cite{Migdal,0022-3700-16-14-006}
(see also \cite{Landau:1991wop}).
Following~\cite{Bernabei:2007jz}, we call these effects the Migdal effects.
In the Migdal's approach, a state of the electron cloud  just after a nuclear recoil 
is approximated by 
\begin{eqnarray}
|\Phi_{ec}'\rangle = e^{- i m_e \sum_{i} {\mathbf v}\cdot \hat{\xe}_i}| \Phi_{ec}\rangle\ ,
\end{eqnarray}
in the rest frame of the nucleus.
Here $m_e$ is the electron mass, 
$\hat\xe_i$ the position operator of the $i$-th electron,
$\mathbf v$ the nucleus velocity after the recoil,
and $|\Phi_{ec}\rangle$ the state of the electron cloud before the nuclear recoil.
The probability of ionization/excitation is then given by
\begin{eqnarray}
\prob = | \langle \Phi_{ec}^* |\Phi_{ec}'\rangle|^2\ ,
\end{eqnarray}
where $|\Phi_{ec}^*\rangle$ denotes either the ionized or excited energy eigenstate of the electron cloud.

In the above analysis, the final state ionization/excitation 
are treated separately from the nuclear recoil.
Thus, the energy-momentum conservation and the probability conservation are made somewhat obscure.
In this paper, we reformulate the Migdal effect so that the ``atomic recoil'' cross section is obtained coherently.
In our reformulation, the energy-momentum conservation and the probability conservation are manifest while the final state 
ionization/excitation are treated automatically.
We also provide numerical estimates of the ionization/excitation probabilities 
for isolated atoms of Ar, Xe, Ge, Na, and I.

The Migdal effect should be distinguished from the ionization and the excitation in scintillation processes.
The Migdal effect takes place even for a scattering of an isolated atom, while the latter occurs due to the interaction between atoms
in the detectors. 
It should be also emphasized that the Migdal effect can ionize/excite electrons in inner orbitals, 
which are not expected in scintillation processes.  
As we will see, the ionization/excitation from the inner orbitals result in extra electronic 
energy injections in the keV range, which can enhance detectability of rather light dark matter in the GeV mass range. 

The organization of the paper is as follows.
In Sec.\,\ref{sec:eigenstates}, we discuss approximate energy eigenstates of an atomic 
state by paying particular attention to the total atomic motion.
In Sec.\,\ref{sec:Migdal}, we reformulate the atomic recoil cross section with the Migdal effect
by taking the energy eigenstates in Sec.\,\ref{sec:eigenstates} as asymptotic states.
In Sec.\,\ref{sec:MESEA}, we calculate the Migdal effect with single electron wave functions.
In Sec.\ref{sec:Numerical}, we estimate the probabilities of the ionization/excitation at a nuclear recoil.
In Sec.\ref{sec:DM}, we discuss implications for dark matter direct detection.
In Sec.\,\ref{sec:neutrino}, we briefly discuss the Migdal effect in a coherent neutrino-nucleus  scattering.
The final section is devoted to our conclusions and discussion.

\section{Energy Eigenstates of Atomic System}
\label{sec:eigenstates}
As we will see in the next section, the plane wave function of a whole atomic system plays a central role to obtain
the nuclear scattering cross section with the Migdal effect. 
In the following, we consider an isolated neutral atom consisting of a nucleus and $N_e$ electrons.
The electrons are not necessarily bounded by the Coulomb potential of the nucleus, 
and hence, the energy eigenstates can be ionic states with unbouded electrons.

As typical nuclear recoil energy is smaller than ${\cal O}(100)$\,keV,
the Hamiltonian of the system is well approximated by the non-relativistic one,
\begin{eqnarray}
\label{eq:hamiltonian}
\hat H_{A} \simeq \frac{\hat {\mathbf p}_N^2}{2m_N} + \hat H_{ec}(\hat{\mathbf x}_N)
=  \frac{\hat {\mathbf p}_N^2}{2m_N} +\sum_i^{N_e}\frac{\hat{\mathbf p}_i^2}{2m_e} +  V(\hat{ \mathbf x}_i - \hat{ \mathbf x}_N)\ .
\end{eqnarray}
Here, $\hat {\mathbf p}_N$ and $\hat{\mathbf x}_N$ denote the momentum and the position
operators of the nucleus with mass $m_N$, respectively.
The momentum and the position operators of the $i$-th electron are given by $\hat {\mathbf p}_i$ and $\hat{\mathbf x}_i$, respectively. 
The Hamiltonian of the electron cloud, $\hat H_{ec}$, depends on the position 
operator of the nucleus, $\hat{\mathbf x}_N$,
through the interaction potential $\hat V(\hat{ \mathbf x}_i - \hat{ \mathbf x}_N)$ ($i = 1 \cdots N_e$).
The interaction potential also includes the interactions between the electrons.
In the coordinate representation, the energy eigen-equation is reduced to
\begin{eqnarray}
\label{eq:totSE}
\left(\frac{\hat {\mathbf p}_N^2}{2m_N} + \hat H_{ec}(\XN)\right) \Psi_E(\XN, \{\xe\}) = E_{A}\,  \Psi_E(\XN, \{\xe\})\ ,
\end{eqnarray}
where the positions (including spinor indices) of the $N_e$ electrons are represented by $\{\xe\}$ collectively.

\subsection{Energy Eigenstates of an Atom at Rest}
To solve Eq.\eqref{eq:totSE}, let us first consider the eigenstates of $\hat H_{ec}(\XN)$ for a given $\XN$,
\begin{eqnarray}
\label{eq:Eec}
\hat H_{ec}(\XN) \Phi_{ec}(\{\xe\}|{\XN}) = E_{ec}(\XN)  \Phi_{ec}(\{\xe\}|\XN)\ .
\end{eqnarray}
Since the system is invariant under spatial translations,
the energy eigenvalues do not depend on $\XN$ while the wave functions depend on $\XN$
only through $\{\xei-\XN\}$;
\begin{eqnarray}
E_{ec}({\XN}) &=& E_{ec} \  , \\
\label{eq:PhiEec}
 \Phi_{E_{ec}}(\{\xe\}|\XN)\  & = &\Phi_{E_{ec}}(\{\xe-\XN\}) \ .
\end{eqnarray}
The eigenstates, $\Phi_{E_{ec}}(\{\xe-\XN\})$, provide a complete orthogonal basis of the electron cloud  
for a given $\XN$.

Next, let us show that $\Phi_{E_{ec}}$ well approximates an energy eigenfunction of the whole atomic system at rest, i.e.,
\begin{eqnarray}
\Psi_{E_A}^{(\rm rest)}(\XN,\{\xe\}) \equiv\Phi_{E_{ec}}(\{\xe-\XN\})\ .
\end{eqnarray}
By substituting $\Psi_{E_A}^{{\rm (rest)}}$ to Eq.\,\eqref{eq:totSE}, the energy eigen-equation results in
\begin{eqnarray}
\label{eq:SE0}
\frac{\hat {\mathbf p}_N^2}{2m_N} \Psi_{E_A}^{{\rm (rest)}}(\XN,\{\xe\})
=  (E_A-E_{ec})
\Psi_{E_A}^{{\rm (rest)}}(\XN,\{\xe\})\ .
\end{eqnarray}
Now, $\Psi_{E_A}^{{\rm (rest)}}$ (i.e. $\Phi_{E_{ec}}$) depends on $\XN$ only through $\{\xe-\XN\}$, 
the momentum of the nucleus is balanced with the electron momentum,
\begin{eqnarray}
{\hat {\mathbf p}_N}\Psi_{E_A}^{{\rm (rest)}}(\XN,\{\xe\})= - \sum_{i}^{N_e} \hat{\mathbf{p}}_i\,
\Psi_{E_A}^{{\rm (rest)}}(\XN,\{\xe\})\ .
\label{eq:rest}
\end{eqnarray}
Thus, the left-hand side of Eq.\,\eqref{eq:SE0} is expected to be highly suppressed, i.e.,
\begin{eqnarray}
\label{eq:kin}
\left\langle\frac{\hat {\mathbf p}_N^2}{2m_N}\right \rangle \sim
\frac{m_e}{m_N} \times E_{ec}\ ,
\end{eqnarray}
for $\Psi_{E_A}^{{\rm (rest)}}$. 
Here, we used the fact that the expectation value of the electron kinetic energy is roughly given by
\begin{eqnarray}
\left\langle\frac{\hat{\mathbf{p}}_i^2}{2m_e} \right\rangle \sim \frac{E_{ec}}{N_e}\ .
\end{eqnarray}
Therefore, we find that $\Psi_{E_A}^{{\rm (rest)}}$ provides  an approximate energy eigenstate of the whole atomic system with 
$E_A \simeq E_{ec}$;
\begin{eqnarray}
\label{eq:EEMA}
\hat H_A \Psi_{E_A}^{{\rm (rest)}}(\XN,\{\xe\}) \simeq E_{ec} \Psi_{E_A}^{{\rm (rest)}}(\XN,\{\xe\})\ .
\end{eqnarray}
It should be noted that this is nothing but the Born-Oppenheimer approximation.%
\footnote{In passing, Eq.\,(\ref{eq:rest}) means that the state $\Psi_{E_A}^{{\rm (rest)}}$ is also an eigenstate of the total momentum
of the atom, i.e.
\begin{eqnarray}
\label{eq:rest2}
\left({\hat {\mathbf p}_N} + \sum_{i=1}^{N_e} \hat{\mathbf p}_i \right) 
\Psi_{E_A}^{{\rm (rest)}}(\XN,\{\xe\})= 0\ .
\end{eqnarray}}

\subsection{Energy Eigenstates of a Moving Atom}
Once we have the energy eigenstates of an atomic system at rest,
the energy eigenstates of a moving atom with a velocity ${\mathbf v}$ can be 
immediately obtained by the Galilei transformation,
\begin{eqnarray}
\Psi_{E_A}(\XN,\{\xe \}) \simeq U(\mathbf v) \Psi_{E_A}^{{\rm (rest)}}(\XN,\{\xe\})\ .
\end{eqnarray}
Here the unitary operator of the Galilei transformation is given by
\begin{eqnarray}
U(\mathbf v) = \exp\left[
i m_N {\mathbf v}\cdot \XN  +  i m_e\sum_{i=1}^{N_e} {\mathbf v}\cdot \xei
\right]\ .
\end{eqnarray}
Under the  Galilei transformation, the momentum operators are shifted by
\begin{eqnarray}
U(\mathbf v)^\dagger \hat{\mathbf p}_N U(\mathbf v) &=& \hat{\mathbf p}_N + m_N \mathbf v \ , \\
U(\mathbf v)^\dagger \hat{\mathbf p}_i U(\mathbf v) &=& \hat{\mathbf p}_i + m_e \mathbf v \ ,
\end{eqnarray}
and the Hamiltonian is transformed into
\begin{eqnarray}
\label{eq:HMA}
U({\mathbf v})^\dagger \hat H_A U(\mathbf v) = \hat H_A
 +  {\mathbf v} \cdot 
 \left(
\hat{\mathbf p}_N
+ \sum_{i=1}^{N_e}\hat{\mathbf p}_i
 \right) 
+\frac{1}{2}\nominalMass  v^2 \ .
\end{eqnarray}
Here we define the nominal mass of the atom by
\begin{eqnarray}
\nominalMass = m_N + N_e m_e \ .
\end{eqnarray}
By using Eqs.\,\eqref{eq:EEMA}, \eqref{eq:rest}, and \eqref{eq:HMA}, we find that the boosted wave function $\Psi_{E_A}$ satisfies,
\begin{eqnarray}
\hat H_A \Psi_{E_A}(\XN,\{\xe \})\simeq \left(E_{ec} + \frac{1}{2}\nominalMass v^2\right)   \Psi_{E_A}(\XN,\{\xe \})\ .
\end{eqnarray}
Therefore,  the boosted wave function $\Psi_{E_A}$ provides the approximate energy eigenstate of a moving atom with energy
\begin{eqnarray}
E_A \simeq E_{ec} + \frac{1}{2}\nominalMass v^2  \ .
\end{eqnarray}

In summary, the eigenstate of the atomic system is approximated by
\begin{eqnarray}
\label{eq:eigenPSI}
\Psi_{E_A}(\XN,\{\xe \}) &\simeq& e^{i {\mathbf p}_N \cdot \XN} 
e^{ i \sum_{i=1}^{N_e}{\mathbf q_e} \cdot \xei }\Psi_{E_A}^{{\rm (rest)}}(\XN,\{\xe\})\ ,\\
{\mathbf p}_N &=&m_N {\mathbf v}\ , \\
{\mathbf q}_e &=&m_e {\mathbf v}\ ,
\end{eqnarray}
with 
\begin{eqnarray}
E_A \simeq E_{ec} + \frac{1}{2}\nominalMass v^2  \ .
\end{eqnarray}
It should be remembered  that $\Psi_{E_A}(\XN,\{\xe\})$ is not an eigenstate of the nucleus momentum $\hat{\mathbf p}_N$.
Instead, $\Psi_{E_A}(\XN,\{\xe\})$ is an eigenstate of the momentum of the whole atom;
\begin{eqnarray}
\left({\hat {\mathbf p}_N} + \sum_{i}^{N_e} \hat{\mathbf p}_i \right) 
\Psi_{E_A}(\XN,\{\xe\}) 
= \left(\nominalMass {\mathbf v}\right)\times \Psi_{E_A}(\XN,\{\xe\}) \ .
\end{eqnarray}
Thus, ${\mathbf p}_N$ in Eq.\,\eqref{eq:eigenPSI} parametrizes not the nucleus momentum but the eigenvalue of the total momentum 
${\mathbf p}_{A} = \nominalMass/m_N \times {\mathbf p}_N = \nominalMass {\mathbf v}$.
It should be also noted that the energy eigenstate in Eq.\,\eqref{eq:eigenPSI} is no more in the realm of the Born-Oppenheimer approximation for $\mathbf v \neq 0$ since they are not eigenfunctions of $\hat H_{ec}$ for a given $\XN$.

\section{MIGDAL EFFECT : From Nuclear Recoil to  Atomic Recoil}
\label{sec:Migdal}
In this section, we derive the recoil cross section of the atomic system with the final state ionization/excitation.
\subsection{Isolated Nuclear Recoil}
\label{sec:NuclearRecoil}
Before proceeding further, let us first translate the dark matter-nucleus interaction in field theory to an interaction potential, which will be 
useful in the later analysis. 
For now, let us forget the electron cloud and take the nucleus as a free separated particle. 
In a relativistic field theory, the $T$-matrix and the invariant amplitude of a scattering process are given by
\begin{eqnarray}
\label{eq:invariantA}
T_{FI} = \langle {\mathbf p}_{N}^F {\mathbf p}_{DM}^F | 
{\mathbf p}_{N}^I {\mathbf p}_{DM}^I \rangle
= {\cal M}\times i(2\pi)^4 
\delta^{4}(
p_{N}^F+p_{DM}^F
-p_{N}^I
-p_{DM}^I
)\ .
\end{eqnarray}
Here, the plane waves of the dark matter and the nucleus are normalized by
\begin{eqnarray}
\label{eq:norm}
\langle {\mathbf p} | {\mathbf p'} \rangle  = (2\pi)^3 2p^0 \delta^3(\mathbf p' - \mathbf p) \ ,
\end{eqnarray}
with $p^0$ being the relativistic energy of the particle. 

As an example, let us consider a contact spin-independent interaction between a Dirac dark matter and nucleons;
\begin{eqnarray}
{\cal L} = \sum_{i=p,n}\frac{g_{i}}{M_*^2} \bar\psi_{i} \psi_i \bar\psi_{DM}\psi_{DM}\ ,
\end{eqnarray}
where $M_*$ denotes a mass parameter and $g_{p,n}$ are dimensionless coupling constants.
In this case, the squared  invariant amplitude for the nucleus scattering is given by
\begin{eqnarray}
|{\cal M}|^2 =  16 \frac{m_{N}^2 m_{DM}^2}{M_*^4} \left(g_p Z + g_n (A-Z)\right)^2\ ,
\end{eqnarray}
where $Z$ is the atomic number,  $A$ the mass number, and $m_{DM}$ the mass of the dark matter.
The corresponding cross section is given by
\begin{eqnarray}
\label{eq:CS0}
\bar\sigma_{N} &\simeq& \frac{1}{16\pi} 
\frac{|{\cal M}|^2}{(m_N + m_{DM})^2} \ ,\\
&\simeq& \frac{1}{\pi} 
\frac{\mu_N^2}{M_*^4} \left(g_p Z + g_n (A-Z)\right)^2\ ,
\end{eqnarray}
where $\mu_N$ is the reduced mass,
\begin{eqnarray}
\mu_N = \frac{m_N m_{DM}}{m_N+m_{DM}}\ .
\end{eqnarray}

In the coordinate representation of quantum mechanics,  the above invariant matrix element in Eq.\,(\ref{eq:invariantA}) is reproduced by an interaction potential,
\begin{eqnarray}
\hat H &=& \hat H_0 + \hat V_{\rm int}\ , \\
\hat H_0&=& \frac{\hat{\mathbf p}_N^2}{2m_N}   + \frac{\hat{\mathbf p}_{DM}^2}{2m_{DM}}   \ , \\
\label{eq:int}
\hat V_{\rm int} &=& \frac{-\cal M}{4m_Nm_{DM}} \delta^3(\XN - \XDM)\ ,
\end{eqnarray}
with the initial and the final states
\begin{eqnarray}
\label{eq:init}
\psi_I(\XN,\XDM) &=& \sqrt{2m_{N}}\, e^{i {\mathbf p}_N^I\cdot \XN} 
\times \sqrt{2m_{DM}}\,e^{i {\mathbf p}_{DM}^I\cdot \XDM} \ ,\\
\label{eq:final}
\psi_F(\XN,\XDM) &=& \sqrt{2m_{N}}\, e^{i {\mathbf p}_N^F\cdot \XN} 
\times \sqrt{2m_{DM}}\,e^{i {\mathbf p}_{DM}^F\cdot \XDM} \ .
\end{eqnarray}
Here, we normalize the initial and the final wavefunctions in conforming with the one in Eq.\,\eqref{eq:norm}
with the relativistic energies approximated by their masses.

{
As another example, we may also consider a dark matter interaction with nucleons
through an exchange of a light scalar particle, $\phi$, with mass $m_{\phi}$,
\begin{eqnarray}
{\cal L} = - \sum_{i=p,n} y_{i} \phi \, \bar\psi_{i} \psi_i - y_{DM} \phi \, \bar\psi_{DM}\psi_{DM}\ ,
\end{eqnarray}
where $y_{p,n,DM}$ are Yukawa coupling constants. 
The invariant amplitude of the isolated nuclear scattering for each spin is given by
\begin{eqnarray}
{\cal M}(q_N^2) &\simeq &y_{DM}\left(y_p Z + y_n (A-Z)\right) \frac{4 m_{DM} m_{N}}{m_{\phi}^{2} - t } \ , \\
t & \simeq & - q_N^2 = - ({\bf p}^{F}_{N} - {\bf p}^{I}_{N})^2 \ ,
\end{eqnarray}
in the non-relativistic limit. 
In the coordinate representation of  quantum mechanics, the invariant amplitude is reproduced by adding a potential term
\begin{eqnarray}
\hat V_{\rm int}({\mathbf x}_N - {\mathbf x}_{DM}) = - \int \frac{d^3{\mathbf q}}{(2\pi)^3}e^{i\mathbf q \cdot ({\mathbf x_N} - {\mathbf x_{DM}})} 
\frac{{\cal M}(q^2)}{4 m_{DM}m_N} \ ,
\end{eqnarray}
with the initial and the final state wave functions in Eqs.\,(\ref{eq:init}) and (\ref{eq:final}).

In both cases, the differential cross section with respect to the nuclear recoil energy in the laboratory frame is given by%
\footnote{The elastic nuclear recoil energy is related to the scattering angle in the center of the mass frame via
\begin{eqnarray}
dE_R = \frac{\mu_N^2}{m_N} v_{DM}^2 \times d\cos\theta_{CM}\ .
\end{eqnarray} } 
\begin{eqnarray}
\label{eq:ds}
\frac{d\sigma_N}{dE_R} &\simeq &\frac{1}{32\pi} 
\frac{m_N}{\mu_N^2 v_{DM}^2} 
\frac{|F_A(q_N^2)|^2|{\cal M}(q_N^2)|^2}{(m_N + m_{DM})^2} 
=\frac{1}{2} 
\frac{m_N}{\mu_N^2 v_{DM}^2} 
\tilde{\sigma}_{N}(q_N)
 \ . 
\label{eq:dssigma}
\end{eqnarray}
Here, we introduce the nuclear form factor, which is relevant for a momentum transfer $q_N$ in the tens to hundreds MeV.
In the last equality, we defined
\begin{eqnarray}
\tilde\sigma_{N}(q_N) = \frac{1}{16\pi}\frac{|F_A(q_N^2)|^2|{\cal M}(q_N^2)|^2}{(m_N + m_{DM})^2}\ ,
\end{eqnarray}
which reduces to $|F_A(q_N^2)|^2\times\bar\sigma_N$ for the contact interaction.
}

\subsection{Invariant Amplitudes with  Electron Cloud}
Now, let us calculate the cross section of the nuclear recoil in the presence of electron cloud.
For this purpose, we consider
\begin{eqnarray}
\hat H_{\rm tot} = \hat H_A+ \frac{\hat{\mathbf p}_{DM}^2}{2m_{DM}}   + \hat V_{\rm int} \ ,
\end{eqnarray}
in the coordinate representation, where $\hat H_A$ is given in Eq.\,\eqref{eq:hamiltonian}.
In subsection\,\ref{sec:NuclearRecoil}, we considered the asymptotic states consist of the plane waves of
dark matter and an isolated nucleus.
To take into account the electron cloud, we replace the plane waves of the nucleus with the plane waves 
of the atomic system discussed in section\,\ref{sec:eigenstates}.

The initial and the final states of dark matter scattering are taken to be
\begin{eqnarray}
\label{eq:states}
\Psi_I(\XN,\{\xe\},\XDM) &=& 
\sqrt{2m_N}\Psi_{E_A^I} (\XN,\{\xe\})\times \sqrt{2m_{DM}} e^{i {\mathbf p}_{DM}^I\cdot \XDM}\ , \\
\Psi_F(\XN,\{\xe\},\XDM) &=& 
\sqrt{2m_N}\Psi_{E_A^F} (\XN,\{\xe\})\times \sqrt{2m_{DM}} e^{i {\mathbf p}_{DM}^F\cdot \XDM} \ .
\end{eqnarray}
Hereafter, we consider the initial atom at rest in the laboratory frame, ${\mathbf v}_I = 0$.
The total energies of the initial and the final states are given by
\begin{eqnarray}
E_I &=& E_{ec}^I + \frac{{\mathbf p}_{DM}^{I}{}^2}{2m_{DM}}\ , \\
E_F &=& E_{ec}^F + \frac{\nominalMass}{2}v_F{}^2+ \frac{{\mathbf p}_{DM}^{F}{}^2}{2m_{DM}}\ ,
\end{eqnarray}
where $E_{ec}^{I,F}$ are the energy eigenvalues of the initial and the final electron clouds 
in the rest frame, respectively.
By using the energy eigenfunctions in Eq.\,\eqref{eq:eigenPSI}, the $T$-matrix of this process is given by
\begin{eqnarray}
i T_{FI} &=& - i(2\pi) \delta(E_F - E_I) \int d^3 \XN d^3 \XDM  \prod_id^3\xei \, 2m_{DM} 2m_{N} \hat{V}_{\rm int}(\XN - \XDM) \cr
&&\times
\Phi_{E_{ec}^F}^*(\{\xe-\XN\}) e^{- i \sum_i {\mathbf q}_e\cdot \xei }e^{-i {\mathbf p}_N^F \cdot \XN}
 \Phi_{E_{ec}^I}(\{\xe-\XN\})e^{-i ({\mathbf p}_{DM}^F -{\mathbf p}_{DM}^I)\cdot \XDM} \ , \\
&= & - i(2\pi)^4 \delta(E_F - E_I)  
\delta^3(\nominalMass {\mathbf v}_F  + {\mathbf p}_{DM}^F  - {\mathbf p}_{DM}^I  ) 
F_A(q_A){\cal M}(q_A) \nonumber\\
&& \times \int \prod_id^3\xei \,\Phi_{E_{ec}^F}^*(\{\xe\}) e^{- i \sum_i {\mathbf q}_e \cdot \xei }
\Phi_{E_{ec}^I}(\{\xe\}) \ . \\
q_A^2 & = & ({\mathbf p}_{DM}^F  - {\mathbf p}_{DM}^I )^2\ .
\end{eqnarray}
In the second equality, we shifted the integration variables $\xei$ and $\XDM$ by $\XN$.

As a result, we obtain the matrix element 
\begin{eqnarray}
i T_{FI} &\simeq& F_A(q_A^2) {\cal M}(q_A^2) \times Z_{FI}(\mathbf q_e) \times
i(2\pi)^4 \delta^4(p_F-p_I) \ ,
\end{eqnarray}
where 
\begin{eqnarray}
\delta^4(p_F-p_I)  &=& \delta\left(E_F-E_I\right)\times \delta^3(\nominalMass{\mathbf v}_F+ {\mathbf p}_{DM}^F  - {\mathbf p}_{DM}^I  )\ ,  \\
Z_{FI}({\mathbf q}_e)
&=&\int \prod_id^3\xei \,\Phi_{E_{ec}^F}^*(\{\xe\}) e^{- i \sum_i{\mathbf q}_e\cdot \xei }\Phi_{E_{ec}^I}(\{\xe\}) \ ,
\label{eq:ZFI}\\
{\mathbf q}_e&=& m_e {\mathbf v}_F\ ,
\end{eqnarray}
for $\mathbf v_I = 0$.%
\footnote{ The Fermi's golden rule is justified by taking the timescale much loner than $(E_{ec}^F-E_{ec}^I)^{-1}$. 
This timescale is also much longer than the typical radius of the electron cloud divided by the speed of light,
and hence, the use of the electrostatic potential is also justified.
}
The term proportional to ${\cal M}$ denotes the nuclear recoil while 
the factor $Z_{FI}(\mathbf q_e)$ denotes the transition of the electron cloud.
It should be emphasized that our approach treats the nucleus and the electron cloud coherently.
This treatment enables us to derive the invariant amplitude with manifest energy-momentum conservation.

\subsection{Phase Space Integration}
By noting the normalizations in Eqs.\,\eqref{eq:eigenPSI},  \eqref{eq:norm}, (see also \eqref{eq:projection0}),
the differential cross section is given by%
\footnote{The factor $|Z_{FI}|^2$ is missing in the cross section in \cite{Vergados:2004bm,*Moustakidis:2005gx,*Ejiri:2005aj,*Vergados:2013raa}. 
}
\begin{eqnarray}
d\sigma &\simeq& 
 \sum_{E_{ec}^F} 
 \frac{d^3{\mathbf p}^F_{A}}{(2\pi)^3 2{p}_{A}^{F}{}^0}
\frac{d^3{\mathbf p}^F_{ DM}}{(2\pi)^3 2p_{DM}^{F}{}^0}
\frac{|F_A(q_A^2)|^2 |{\cal M}(q_A^2)|^2
\times |Z_{FI}({\mathbf q}_e)|^2}{4 \sqrt{(p_{A}^I\cdot p_{DM}^I)^2 - \physicalMass^2 m_{DM}^2}}
\nonumber\\
&&\hspace{5cm}\times (2\pi)^4 \delta^4(p_{A}^F + p_{DM}^F  - p_{A}^I - p_{DM}^I) \ .
\end{eqnarray}
Here, we defined the physical mass of the atomic system, $\physicalMass$ by
\begin{eqnarray}
\physicalMass = \nominalMass + E_{ec}\ .
\end{eqnarray}
By boosting four momentum $(\physicalMass,0,0,0)$, we obtain the four-momentum of the atomic system in an arbitrary frame.
For example, the final state four-momentum is given by
\begin{eqnarray}
p_{A}^F \simeq (p_{A}^F{}^0, \nominalMass {\mathbf v}_F  )\ , 
\quad p_{A}^F{}^0 \simeq \physicalMass^F + \frac{1}{2}\nominalMass v_F^2
= \nominalMass + E_{ec}^F + \frac{1}{2}\nominalMass v_F^2\ ,
\end{eqnarray}
in the laboratory frame.

{
When the magnetic quantum numbers of the electrons in the initial/final states are averaged/summed, 
the factor $|Z_{FI}({\mathbf q}_e)|^2$ depends only on the size of ${\mathbf q}_e$.
In this case, the differential cross section is given by
\begin{eqnarray}
\frac{d\sigma}{d\cos\theta_{CM} } &\simeq& 
\sum_{E_{ec}^F} \frac{1}{32\pi}
\frac{|{\mathbf p}_F|}
{(p_{A}^{I}{}^0 + p_{DM}^{I}{}^0)^2 |{\mathbf p}_I|}
|F_A(q_A^2)|^2 |{\cal M}(q_A^2)|^2
 |Z_{FI}(q_e)|^2 \ .
\end{eqnarray}}
Here, ${\mathbf p}_{I,F}$ denotes the initial and the final state momenta in the center of the mass frame.

By using the dark matter velocity in the laboratory frame, ${\mathbf v}_{DM}^I$, 
the initial momentum in the center of the mass frame, ${\mathbf p}_I$, is given by
\begin{eqnarray}
{\mathbf p}^I_{DM} = - {\mathbf p}^I_A = {\mathbf p}_I \simeq  \mu_N {\mathbf v}_{DM}^I \ .
\end{eqnarray}
It should be noted that the scattering process is no longer elastic for $E_{ec}^F \neq E_{ec}^I$.
Accordingly, the final state momentum in the center of the mass frame becomes smaller than $|{\mathbf p}_I|$;
\begin{eqnarray}
\label{eq:pF}
|{\mathbf p}_F|^2 &\simeq& |{\mathbf p}_I|^2 - 2 \mu_N (E_{ec}^F - E_{ec}^I) \ .
\end{eqnarray}
To satisfy $|{\mathbf p}_F| > 0$, there is a threshold velocity, 
\begin{eqnarray}
\label{eq:vDMth}
v_{DM}^{(th)} =\sqrt{  \frac{2(E_{ec}^F - E_{ec}^I)}{\mu_N}}\ ,
\end{eqnarray}
with which  $|{\mathbf p}_{F}|$ is rewritten by
\begin{eqnarray}
|{\mathbf p}_{F}| = \mu_N \sqrt{v_{DM}^2 - v_{DM}^{(th)\, 2} }\ .
\end{eqnarray}

\subsection{Atomic Recoil Spectrum}
The atomic recoil spectrum in the laboratory frame is obtained as follows.%
\footnote{Similar kinematics has been discussed in the context of 
``inelastic excitation of nucleus'' in \cite{Ellis:1988nb,Bernabei:2000qn}.}
The atomic recoil energy $E_R$ in the laboratory frame is given by
\begin{eqnarray}
E_R = p_{A}^{F}{}^0 - \physicalMass^F \simeq \frac{1}{2} \nominalMass v_F^2\ .
\end{eqnarray}
As the momentum transfer is given by
\begin{eqnarray}
q_A^2 &\simeq&  (|{\mathbf p}_F| - |{\mathbf p}_I |)^2 
+ 2 |{\mathbf p}_I ||{\mathbf p}_F | (1 - \cos\theta_{CM})\ ,\\
&\simeq& - (E_{ec}^F - E_{ec}^I)^2+ 2 \physicalMass E_R \simeq 2 \physicalMass E_R\ ,
\end{eqnarray}
we obtain 
\begin{eqnarray}
E_R &\simeq& \frac{q_A^2}{2\physicalMass} 
\simeq 
\frac{|{\mathbf p}_F|^2 + |{\mathbf p}_I |^2 - 2 |{\mathbf p}_I ||{\mathbf p}_F | \cos\theta_{CM}}{2\physicalMass}\ .
\end{eqnarray}
Thus, the differential cross section with respect to the atomic recoil energy is given by 
\begin{eqnarray}
\frac{d\sigma}{dE_R} &\simeq& 
\sum_{E_{ec}^F} 
\frac{1}{32\pi}
\frac{m_A}{\mu_N^2 v_{DM}^2} 
\frac{|F_A(q_A^2)|^2|{\cal M}(q_A)|^2}{(m_A + m_{DM})^2} 
 |Z_{FI}(q_e)|^2 \ , \\
  \label{eq:constM}
 &\simeq& 
\sum_{E_{ec}^F} 
\frac{1}{2} 
\frac{m_A}{\mu_N^2 v_{DM}^2} 
\tilde{\sigma}_N(q_A)
 |Z_{FI}(q_e)|^2 \ , 
\end{eqnarray}
where 
\begin{eqnarray}
q_e = m_e v_{F}\simeq \frac{m_e}{\physicalMass} q_A \ .
\end{eqnarray}

Finally, the dark matter event rate for unit detector mass is given by
\begin{eqnarray}
\frac{dR}{dE_{R} dv_{DM}} 
&\simeq&\frac{1}{m_A} \frac{\rho_{DM}}{m_{DM}}  \frac{d \sigma}{dE_{R}} v_{DM} \tilde f_{DM}(v_{DM}) \ ,\\
&\simeq& \sum_{E_{ec}^F} 
\frac{1}{2}
\frac{\rho_{DM}}{m_{DM}}
\frac{1}{\mu_N^2 }
\tilde{\sigma}_N(q_A)
\times 
|Z_{FI}(q_e)|^2 
 \times 
 \frac{\tilde f(v_{DM})}{v_{DM}}\  .
 \label{eq:recoilspec}
\end{eqnarray}
Here, $\rho_{DM}$  denotes 
the local dark matter density%
\footnote{For the Burkert profile~\cite{Nesti:2013uwa}, for example,
it is estimated to be $\rho_{DM}\simeq 0.487^{+0.075}_{-0.088}$\,GeV/cm$^3$.}
and {$\tilde f(v_{DM})$ is the dark matter velocity distribution integrated over the directional component}
normalized by%
\footnote{For astrophysical uncertainties of the direct detection experiments (see e.g.~\cite{McCabe:2010zh,Green:2011bv}).}
\begin{eqnarray}
\int \tilde{f}_{DM}(v_{DM}) \,dv_{DM} = 1 \ .
\end{eqnarray}

\section{Migdal Effect In  Single Electron Approximation}
\label{sec:MESEA}
\subsection{Single Electron Wave Function}
\label{sec:SEA}
In our numerical calculation, we use the electron wave function, $\Phi_{E_{ec}}$,
obtained by the Dirac-Hartree-Fock method, where 
the relativistic effects on the electron cloud are taken into account (see e.g.~\cite{9783540680109} for review).

In the Dirac-Hartree-Fock approximation, an electron state is given by a Slater determinant made up of one  orbital per each electron
in an averaged central potential around a nucleus.
In this approximation, the energy eigenstates in Eq.\,\eqref{eq:eigenPSI} are approximated by
\begin{eqnarray}
\label{eq:Slater}
\Psi_{E_A}(\XN,\{\xe\}) &\simeq&   e^{i {\mathbf p}_N \cdot \XN} 
\sum_{\sigma \in S_{N_e}}  \frac{\rm sgn(\sigma)}{\sqrt{N_e!}}
e^{ i {\mathbf q}_e \cdot \xe_1 }
 \phi^{\a_1}_{{o}_{\sigma(1)}}(\xe_1-\XN)
e^{ i {\mathbf q}_e \cdot \xe_2 }  \phi^{\a_2}_{{o}_{\sigma(2)}}(\xe_2-\XN) \nonumber\\
&& \hspace{3cm} \times \cdots
e^{ i {\mathbf q}_e \cdot \xe_{N_e} }    \phi^{\a_{N_e}}_{{o}_{\sigma(N_e)}}(\xe_{N_e}-\XN)\ ,
\end{eqnarray}
where {$S_{N_e}$ denotes the permutation group of degree $N_e$}.
Here we explicitly show the indices of the Dirac spinor by $\alpha_i= 1\cdots 4$, which are 
encapsulated in $\{\xe\}$ on the left-hand side.

The electron cloud consists of the $N_e$ orbitals,%
\footnote{The Slater determinant in Eq.\,\eqref{eq:Slater} is reducible in terms of the total 
angular momentum of the atom.}
\begin{eqnarray}
ec = \{o_1, o_2, \cdots o_{N_e}\}\  ,
\end{eqnarray}
where each orbital is specified by  energy $E$,  relativistic angular momentum $\k$, and  magnetic quantum number  $m$,%
\footnote{The value $\k$ determines both the total angular momentum $j$ and the orbital angular momentum $\ell$
via $\k = \mp(j+1/2) $ for $j = \ell \pm 1/2$.}
\begin{eqnarray}
o_i = (E_i, \kappa_i, m_i)\ .
\end{eqnarray}
For a bounded electron, i.e. $E_i<0$, the state is classified by the principle quantum number, $n_i$, 
while the spectrum is continuous for an unbounded electron, i.e. for $E_i > 0$.

The one electron Dirac orbital $\phi_o^{\a}(\xei)$ 
is given by using the {two-component} spherical spinors $\Omega_{\k m}$; 
\begin{eqnarray}
\phi_{o}(\xe)  = \frac{1}{r} 
\left(
\begin{array}{ccc}
 P_{E}(r) \Omega_{\k m}(\theta,\varphi)  \\ 
iQ_{E}(r) \Omega_{-\k m}(\theta,\varphi)
\end{array}
\right)\ .
\end{eqnarray}
{See e.g.~\cite{9783540680109} for the details of the spherical spinors and 
the radial wave functions, $P_{E}(r)$ and $Q_{E}(r)$}.
Here, an atom is at rest and $r$ denotes the distance between the electron and the center of the potential.
The one-electron states are normalized such that
\begin{eqnarray}
\sum_{\a=1}^{4}\int d^3\xe \,\phi_{o}(\xe)^{\a*} \phi_{o'}^\a(\xe) = 
\begin{cases}
\delta_{nn'}\delta_{\k\k'} \delta_{mm'}  & ({\rm bounded})\\
(2\pi)\delta(E-E')\delta_{\k\k'} \delta_{mm'}  &({\rm unbounded} )
\end{cases}
\  .
\end{eqnarray}

In the Dirac-Hartree-Fock approximation,
the electron cloud transition factor in Eq.\,(\ref{eq:ZFI}) is
rewritten by
\begin{eqnarray}
Z_{FI}({\mathbf q}_e)
= \sum_{\sigma \in S_{N_e}} 
{\rm sgn}(\sigma) \prod_{i=1}^{N_e}
 \sum_{\alpha_i=1}^4 
 \int d^3\xei 
\,\phi_{o_{\sigma(i)}^F }^{\a_i*}(\xei)
e^{- i {\mathbf q}_e\cdot \xei }
\phi_{o_i^I}^{\a_i}(\xei)\ .
\end{eqnarray}
In this approximation, the transition amplitude is given by the product of the transition amplitudes between the electron orbitals.

\subsection{Single Electron Excitation/Ionization}
\label{sec:SEE}
For an atomic recoil with a momentum transfer smaller than the hundreds MeV range, 
the factor $|{\mathbf q_e}\cdot \xei |$ is expected to be small than ${\cal O}(1)$ on the atomic scale.%
\footnote{For $v_F \simeq 10^{-3}$, for example, $q_e \simeq 0.5$\,keV and hence $|{\mathbf q}_e\cdot \xei| \ll 1$ even for
a Bohr radius.}
Thus, we consider the Migdal effect at the leading order of $q_e$.
At the leading order of $q_e$, only one electron can be excited/ionized,
and hence, the  initial and the final state configurations are
\begin{eqnarray}
ce_I &=& \{o_1,\cdots , o_k , \cdots o_{N_e}\} \ , \\
ce_F &=& \{o_1,\cdots , o_k' , \cdots o_{N_e}\} \ ,
\end{eqnarray}
where 
\begin{eqnarray}
E_{ec}^F - E_{ec}^I \simeq E_{k}' - E_k \ .
\end{eqnarray}
Hereafter, we assume that the initial electron cloud stays in the ground state, where 
all the electrons are bounded by the Coulomb potential of the nucleus.
In the final electron state, $o_k'$ can be either a bounded or an unbounded orbital.

At the leading order of $q_e$, the electron cloud transition amplitude is reduced to
\begin{eqnarray}
Z_{FI}({\mathbf q}_e)
&=& z_{\mathbf q_e}(E_k',\k_k',m_k'| E_k,\k_k,m_k)= -i
\sum_{\alpha_k=1}^4   \int 
d^3\xe_k \,
\phi_{o_{k}'}^{\a_k*}(\xe_k)
( {\mathbf q}_e\cdot \xe_k )
\phi_{o_k}^{\a_k}(\xe_k)\ .
\end{eqnarray}
At this order, electron transitions are allowed only when the orbital angular momenta of $o_k'$ and $o_k$ differ by one, i.e. $|\ell_k'-\ell_k| = 1$.
Thus, the transition amplitude is reduced to%
\begin{eqnarray}
z_{\mathbf q_e}(E_k',\k_k',m_k'| E_k,\k_k,m_k) &=& 
-i q_e \int dr\,  r \times \left[ 
P_{E_k'}(r)P_{E_k}(r)
+ Q_{E_k'}(r)Q_{E_k}(r)
\right] \nonumber \\
&&
{
\times \int d\Omega 
\,\Omega^\dagger_{\k',m'} (\theta,\varphi)
\cos\theta\,
\Omega^\dagger_{\k,m} (\theta,\varphi)\ }.
\end{eqnarray}
Here, we take the quantization axis of the angular momentum corresponds to ${\mathbf q}_{e}$.
The angle, $\theta$, is the one between ${\mathbf q}_e$ and ${\mathbf x}_e$. 
The choice of the quantization is irrelevant for the final results as we take an average/sum the magnetic quantum numbers of the electrons in the initial/final states.

In the following discussion, we only require an accuracy of ${\cal O}(10)$\% for the electron binding energies.
For this accuracy, the bound state energies for given principal number and the orbital angular momentum
are not distinguishable, and hence, it is useful to label the bound states by $(n,\ell)$.
Accordingly, the transition rates are also labeled by $(n,\ell)$,
\begin{eqnarray}
\label{eq:transit}
 \sum_{F}|Z_{FI}|^2&=&|Z_{II}|^2+\sum_{n,\ell,n',\ell'}
 p_{q_e}^d(n\ell\to n'\ell')+\sum_{n,\ell}\int\frac{dE_e}{2\pi}\frac{d}{dE_e}p_{q_e}^c(n\ell\to E_e)\ .
\end{eqnarray}
Here, $|Z_{II}|^2 \simeq 1 + {\cal O}(q_e^2 \vev{r}^2)$  is the probability for the electrons unaffected 
by the nuclear recoil (see also  appendix\,\ref{sec:probability} for discussion of the probability conservation). 
The excitation and the ionization probabilities, $p^{d}_{q_e}$ and $p^{c}_{q_e}$, are defined by
\begin{eqnarray}
 p_{q_e}^d(n\ell\to n'\ell')&=&\frac{\omega_{\ell'}^{\max}- \omega_{n',\ell'}}{ \omega_{\ell'}^{\max}}\frac{\omega_{n,\ell}}{\omega_{\ell}^{\max}}\sum_{\kappa,\kappa',m,m'}\delta_{\ell,|\kappa+1/2|-1/2}\delta_{\ell',|\kappa'+1/2|-1/2}
 \label{eq:excitationELL}
 \nonumber\\
&&\times\left|z_{\mathbf q_e}(E_{n'\kappa'},\kappa',m'|E_{n\kappa},\kappa,m)\right|^2\, ,\\
\frac{d}{dE_e} p_{q_e}^c(n\ell\to E_e)&=&\frac{\omega_{n,\ell}}{\omega_{\ell}^{\max}}\sum_{\kappa,\kappa',m,m'}\delta_{\ell,|\kappa+1/2|-1/2}\left|z_{\mathbf q_e}(E_e,\kappa',m'|E_{n\kappa},\kappa,m)\right|^2\, .
 \label{eq:ionizationELL}
\end{eqnarray}
Here, $E_{n\kappa}$ is the size of the binding energy for the bounded electron labeled by ($n,\kappa$), 
$E_e$ the energy of the unbounded electron, $\omega_{n\ell}$ the occupation number of the subshell (see Tab.\,\ref{tbl:ground_cfg}), 
and $\omega_{\ell}^{\max}=2(2\ell+1)$.
The final state orbital angular momentum, i.e. $\ell' = \ell \pm 1$, is summed implicitly in Eq.\,\eqref{eq:ionizationELL}.

\begin{table}[t]
 \begin{tabular}{|c|cccccccccccc|}
 \hline
  &$1s$ &$2s$ &$2p$ &$3s$ &$3p$ &$3d$ &$4s$ &$4p$ &$4d$ &$4f$ &$5s$ &$5p$ \\
\hline
  Na&2 &2 &6 &1 &0 &0 &0 &0 &0 &0 &0 &0 \\
\hline
  Ar&2 &2 &6 &2 &6 &0 &0 &0 &0 &0 &0 &0 \\
  \hline
  Ge&2 &2 &6 &2 &6 &10 &2 &2 &0 &0 &0 &0 \\
  \hline
  I&2 &2 &6 &2 &6 &10 &2 &6 &10 &0 &2 &5 \\
  \hline
  Xe&2 &2 &6 &2 &6 &10 &2 &6 &10 &0 &2 &6\\
  \hline
 \end{tabular}
\caption{The number of electrons in a shell for the ground state configurations.}
\label{tbl:ground_cfg}
\end{table}

\subsection{Ionization Spectrum at the Leading Order}
\begin{figure}[tb]
\begin{minipage}{.45\linewidth}
\begin{center}
  \includegraphics[width=.8\linewidth]{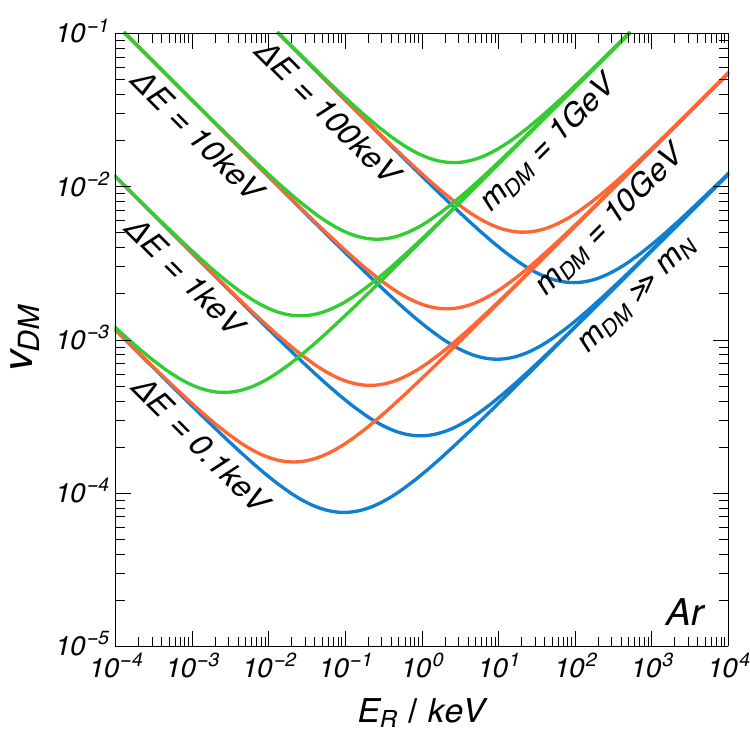}
  \end{center}
  \end{minipage}
 \begin{minipage}{.45\linewidth}
 \begin{center}
  \includegraphics[width=.8\linewidth]{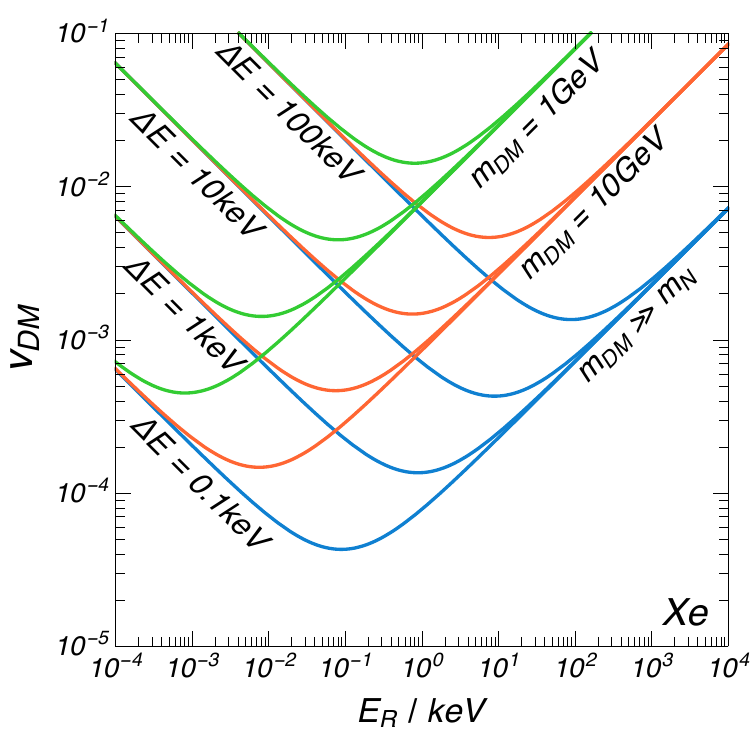}
   \end{center}
  \end{minipage}
\caption{\sl \small
The minimum dark matter velocity as a function of the atomic recoil energy $E_R$ for given ${\mit\D}E$ and $m_{DM}$
for Ar and Xe.}
\label{fig:phase}
\end{figure}

\begin{figure}[tb]
\begin{minipage}{.45\linewidth}
\begin{center}
  \includegraphics[width=.8\linewidth]{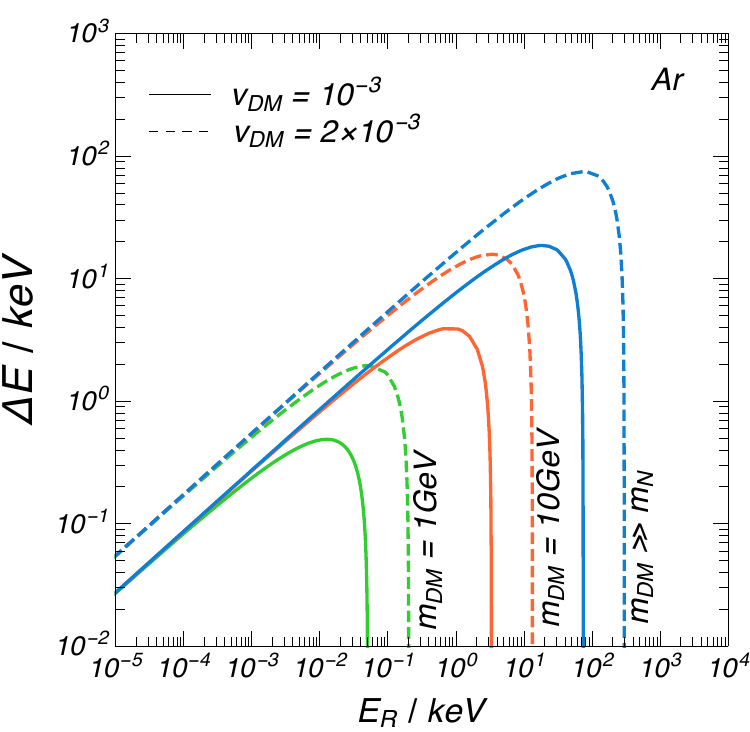}
  \end{center}
  \end{minipage}
 \begin{minipage}{.45\linewidth}
 \begin{center}
  \includegraphics[width=.8\linewidth]{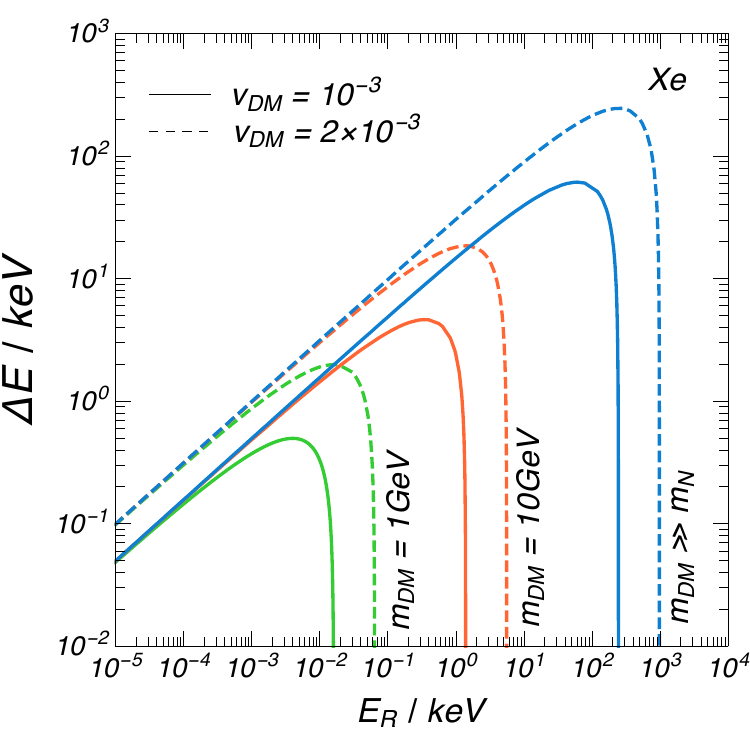}
   \end{center}
  \end{minipage}
\caption{\sl \small
Kinematical constraints on the plane of $(E_R, {\mit \D}E)$ for given $m_{DM}$ and $v_{DM}$ for Ar and Xe atoms.
The regions  below the lines are kinematically allowed.}
\label{fig:kin}
\end{figure}

By combining Eqs.\,\eqref{eq:recoilspec} and \eqref{eq:transit}, we find that the ionized electron spectrum 
from an initial orbital $o_k$ associated is given by
\begin{eqnarray}
\label{eq:ionization}
\frac{dR}{dE_{R}\, dE_e\, dv_{DM}} 
&\simeq&
\frac{dR_0}{dE_{R}\, dv_{DM}} 
 \times \frac{1}{2\pi} \sum_{n,\ell}\frac{d}{dE_e} p^c_{q_e}(n\ell\to E_e)
\ , \\
\frac{dR_0}{dE_{R}\, dv_{DM}} & \simeq & 
\frac{1}{2}
\frac{\rho_{DM}}{m_{DM}}
\frac{1}{\mu_N^2 }
\tilde{\sigma}_N(q_A)
 \times \frac{\tilde f(v_{DM})}{v_{DM}}\  .
\label{eq:electroncont}
\end{eqnarray}
Here, 
\begin{eqnarray}
E_R \simeq \frac{q_A^2}{2m_A} \ , \quad q_e \simeq \frac{m_e}{\physicalMass} q_A \ .
\end{eqnarray}
It should be noted that the atomic recoil energy, $E_R$, and the electron transition energy,  
${\mit \D}E$, are correlated through the energy-momentum conservation;
\begin{eqnarray}
E_R &=& \frac{\mu_N^2}{2m_N} v_{DM}^2 \left( 
\left(1-\sqrt{1 - \frac{2{\mit \D}E}{\mu_N v_{DM}^2}}
\right)^2
+2 (1-\cos\theta_{CM})\sqrt{1 - \frac{2{\mit \D}E}{\mu_N v_{DM}^2}}
\right)\  , 
\end{eqnarray}
where 
\begin{eqnarray}
{\mit \D}E & = & E_{e} + E_{n\ell}\ , \\
E_{n\ell} &=& \frac{1}{2}\sum_\kappa\delta_{\ell,|\kappa+1/2|-1/2}E_{n\kappa} \  .
\end{eqnarray}
From this expression, we find the minimum dark matter velocity for given $E_R$ and ${\mit \D}E$,
\begin{equation}
 v_{DM,\min}\simeq \frac{m_N E_R+\mu_N\mit \D E }{\mu_N\,\sqrt{2\,m_N E_R}}\,.
\end{equation}
In Fig.\,\ref{fig:phase}, we show the minimum velocity as a function of $E_R$ for isolated Ar and Xe atoms.
We also show the kinematically allowed region of $E_R$ and ${\mit \D}E$ for Ar and Xe atoms
for $v_{DM} = 10^{-3}$ in Fig.\,\ref{fig:kin}.

\begin{figure}[tb]
\begin{minipage}{.49\linewidth}
\begin{center}
  \includegraphics[width=.8\linewidth]{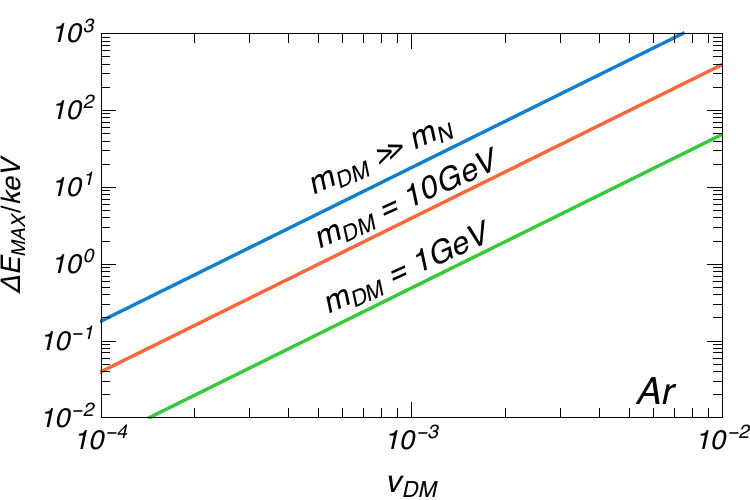}
  \end{center}
  \end{minipage}
 \begin{minipage}{.49\linewidth}
 \begin{center}
  \includegraphics[width=.8\linewidth]{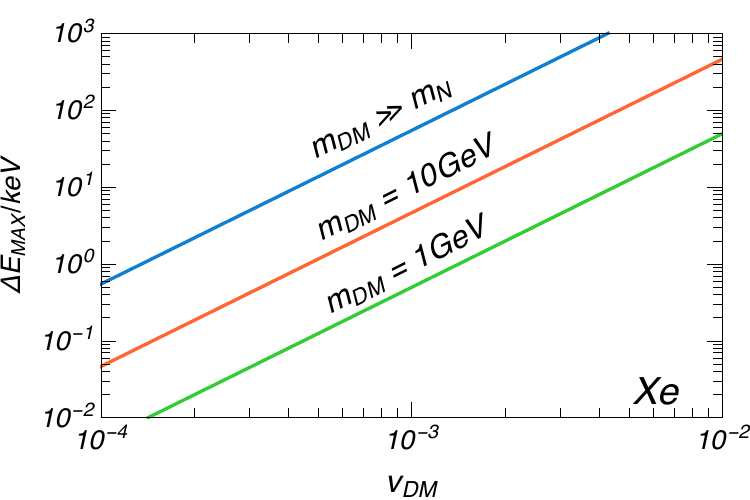}
   \end{center}
  \end{minipage}
\caption{\sl \small
The kinematical upper limits on the electron transition energy, ${\mit\D}E$, 
as a function of the speed of  dark matter for a given $m_{DM}$.
In the solar rest frame,  typical dark matter velocities are of $500$\,km/s with an upper limit of around $700$\,km/s.
}
\label{fig:dEmax}
\end{figure}

It should be also noted that  there is a kinematical upper limit on the electron transition energy, ${\mit\D}E$,
for a given speed of dark matter, which is set by Eq.\,\eqref{eq:vDMth},
\begin{eqnarray}
{\mit \D}E_{\rm MAX} = \frac{1}{2}\mu_N v_{DM}^2 \ .
\end{eqnarray}
In Fig.\,\ref{fig:dEmax}, we show ${\mit \D}E_{\rm MAX}$ as a function of $v_{DM}$.
The figure shows that ${\mit\D}E$ in the keV range is kinematically allowed for $v_{DM}\gtrsim 10^{-3}$.
It is also notable that, for ${\mit \D}E = {\mit \D}E_{\rm MAX} $, the atomic recoil energy is given by
\begin{eqnarray}
 E_R = \frac{\mu_N^2}{2m_N} v_{DM}^2 = \frac{\mu_N}{m_N}\times {\mit \D}E_{\rm MAX}\ .
\end{eqnarray}
Thus the corresponding atomic recoil energy is smaller than ${\mit \D}E_{\rm MAX}$,
which plays an important role on the dark matter detections as discussed  in the later section.

When electrons are emitted from inner orbitals, the created core-holes are de-excited subsequently.%
\footnote{The de-excitation proceeds through the X-ray transition, the Auger transition, 
or the Coster-Kronig transition (see~\cite{BAMBYNEK:1972zz,CAMPBELL2001} for review, see also \cite{2012PhRvA..85f3415S}).
For a core-hole in $n > 1$ states the Coster-Kroning transition dominates the de-excitation process.}
The typical timescales of the de-excitation processes are of ${\cal O}(10)$\,fs. 
Thus, the energies of the electron emission and the de-excitation are measured simultaneously,
and hence, the total electronic energy released at the ionization is given by
\begin{eqnarray}
\label{eq:EEM}
E_{EM} = E_{e} + E_{\rm dex}\ ,
\end{eqnarray}
where $E_{\rm dex}$ is the energy released at the de-excitation.

Accordingly, the electromagnetic energy spectrum is given by
\begin{eqnarray}
\label{eq:deex2}
\frac{dR}{dE_{R}\, dE_{EM} \, dv_{DM}} 
&\simeq& \frac{dR_0}{dE_{R}\, dv_{DM}} 
 \times \frac{1}{2\pi}\sum_{n,\ell}\frac{d}{dE_e} p^c_{q_e}(n\ell\to (E_{EM} - E_{n\ell})) \ .
\label{eq:PhotonFromion}
\end{eqnarray}
Hereafter, we simply assume that the ionization energy is released completely, that is 
  $E_{EM} = {\mit\Delta}E$.%
\footnote{If the atom is completely isolated, the ionization and the subsequent Auger and Coster-Kronig transitions
leave ionized atoms. In the medium, on the contrary, ionized atoms are also de-excited eventually.}
{One caveat here is that $E_{EM}$ is not the energy of a single electron/photon 
but the collection of the energies of the electrons and photons emitted at the de-excitation and the ionization.
Thus, the detector responses to $E_{EM}$ might be different from those to a single electron/photon
with the same energy, although we do not take such effects into account in the following discussion.}

Similarly, the excited atoms also lead to electronic energy release by de-excitation.
Assuming the complete de-excitation again, we obtain the electromagnetic energy spectrum;
\begin{eqnarray}
\label{eq:deex1}
\frac{dR}{dE_{R}\, dE_{EM}\, dv_{DM}} 
\!\!&\simeq&\!\! 
\frac{dR_0}{dE_{R}\, dv_{DM}}
 \times\! \sum_{n,n',\ell,\ell'}p^d_{q_e}(n\ell\to n'\ell') \times \delta(E_{EM} - \mit \D E_{n\ell\to n'\ell'})\ . 
\end{eqnarray}
Here $\mit \D E_{n\ell\to n'\ell'}$ is given by
\begin{equation}
{ \mit \D} E_{n\ell\to n'\ell'}=
 \frac{1}{2}\sum_\kappa\delta_{\ell,|\kappa+1/2|-1/2}E_{n\kappa}-\frac{1}{2}\sum_{\kappa'}\delta_{\ell',|\kappa'+1/2|-1/2}E_{n'\kappa'}\ . 
\end{equation}

\section{Numerical Analysis}
\label{sec:Numerical}
In this section, we provide numerical estimates of the electron transition probabilities,
$p_q^d$ and $p_q^c$, for isolated Ar, Ge, Xe, Na, and I atoms.
To calculate the electron wave functions, we use the {\tt Flexible Atomic Code (FAC, cFAC)}~\cite{Gu2008}.
It is a multi-configuration Dirac-Fock (MCDF) program to calculate various atomic radiative and collisional processes.
We give a brief review of the Dirac-Hartree-Fock method in  appendix\,\ref{sec:DHF}. 
In {\tt FAC}, all of the single electron wave functions, including those of excited and unboudend electrons,
are calculated from a universal central potential, 
\begin{eqnarray}
V({\xe-\XN})&\simeq&  V_{N}(r) + V_{ee}(r)\ ,\\
V_{ee}(r)&=&\frac{\alpha\sum_{n,\kappa}\omega_{n\kappa}\rho_{n\kappa}(r){\cal Q}^{\rm eff}_{n\kappa}(r)}{r\sum_{n,\kappa}\omega_{n\kappa}\rho_{n\kappa}(r)}\ ,
\label{eq:FACpotential}
\end{eqnarray}
which is optimized for the (possible) ground state configurations.
Here,  $V_N(r)$ denotes the Coulomb potential from the nucleus, and  
$\alpha$ the fine-structure constant. 
The factor ${\cal Q}_{n\kappa}^{\rm eff}$ provides the effective charge of the central potential 
for the electrons in the $n\kappa$-orbital;
\begin{align}
 {\cal Q}^{\rm eff}_{n\kappa}(r)&=\sum_{n',\kappa'}\omega_{n'\kappa'}Y_{n'\kappa'}^0(r)-Y_{n\kappa}^0(r)\nonumber\\
 &\hspace{3ex}-(\omega_{n\kappa}-1)\sum_{k>0}f_k(\kappa)Y_{n\kappa}^k(r)\nonumber \\
 &\hspace{3ex}-\sum_{n'\kappa'\neq n\kappa}\frac{\rho_{n\kappa,n'\kappa'}\omega_{n\kappa}\omega_{n'\kappa'}}{\rho_{n\kappa}\omega_{n\kappa}}\sum_kg_k(\kappa,\kappa')Y_{n\kappa,n'\kappa'}^k(r)\, .\label{eq:effective_charge}
\end{align}
Here, $\omega_{n\kappa}$ is the occupation number of the subshell, $\rho_{n\kappa}=\rho_{n\kappa,n\kappa},~
 Y^k_{n\kappa}=Y^k_{n\kappa,n\kappa}$, and
\begin{eqnarray}
\rho_{n\kappa,n'\kappa'}(r)&=&P_{n\kappa}(r)P_{n'\kappa'}(r)+Q_{n\kappa}(r)Q_{n'\kappa'}(r)\, ,\\
 Y_{n\kappa,n'\kappa'}^k(r)&=&r\int\frac{r_<^{k}}{r_>^{k+1}}\rho_{n\kappa,n'\kappa'}(r')dr'\, ,\\
 f_k(\kappa)&=&\left(1+\frac{1}{2j_{\kappa}}\right)
{\small \tj{j_\kappa}{k}{j_{\kappa}}{-1/2}{0}{1/2}}^2\ ,\\
 g_k(\kappa,\kappa')&=&
{\small \tj{j_\kappa}{k}{j_{\kappa'}}{-1/2}{0}{1/2}}^2\ , 
\end{eqnarray}
where $j_{\kappa}$ is the value of $j$ corresponding to $\kappa$, $r_>=\max(r,r')$, $r_<=\min(r,r')$, and 
{\tiny$\tj{j_1}{j_2}{j_3}{m_1}{m_2}{m_3}$} is the Wigner $3j$ symbol.

Due to the universal central potential, the wave functions of orbitals obtained by 
{\tt FAC} are orthogonal to each other, and hence, they can be used to calculate the 
transition rates  in the previous sections.
In our analysis, we further approximate the atomic state by a single Slater determinant of the wave functions 
labeled by a set of $\{n,\kappa,m\}$.
The energy eigenvalues obtained with this approximation reproduce the measured values in Ref.\,\cite{thompson2001x} 
at the accuracy better than $20$\%, which is good enough for later discussion (see the energy levels in 
Tab.\,\ref{tbl:excitation}.).

\subsection{Transition probabilities}

\begin{figure}[tb]
\begin{minipage}{.49\linewidth}
\begin{center}
  \includegraphics[width=.9\linewidth]{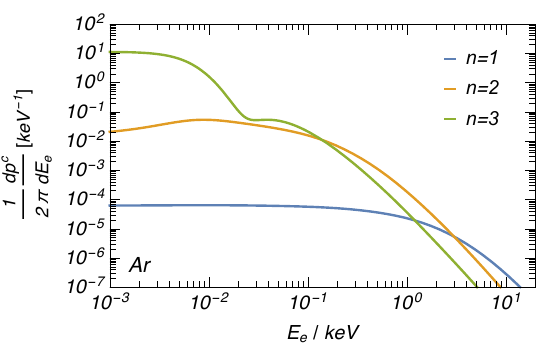}
  \end{center}
  \end{minipage}
 \begin{minipage}{.49\linewidth}
 \begin{center}
  \includegraphics[width=.9\linewidth]{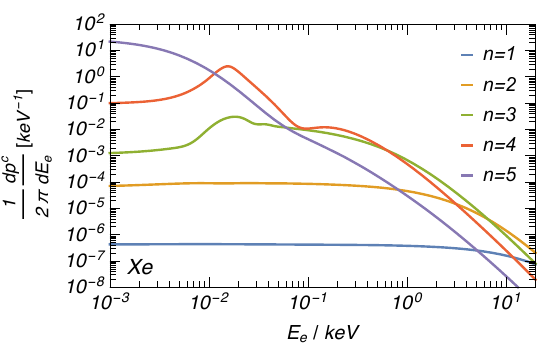}
   \end{center}
  \end{minipage}
  \begin{minipage}{.49\linewidth}
\begin{center}
  \includegraphics[width=.9\linewidth]{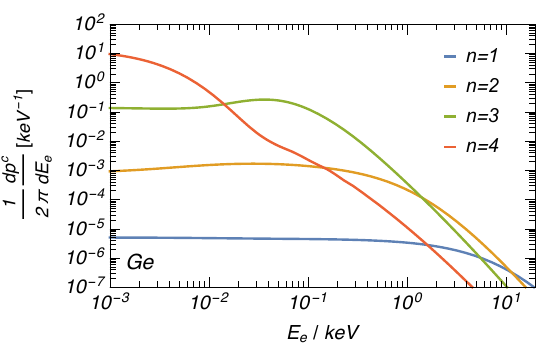}
  \end{center}
  \end{minipage}
 \begin{minipage}{.49\linewidth}
 \begin{center}
  \includegraphics[width=.9\linewidth]{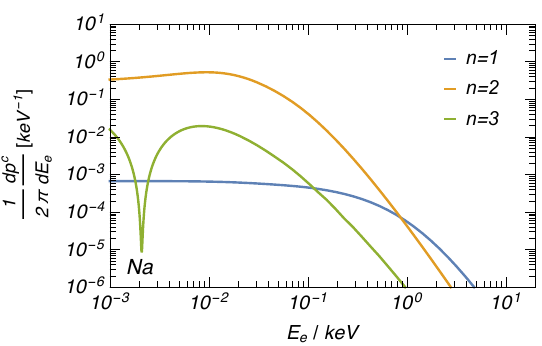}
   \end{center}
    \end{minipage}
   \begin{minipage}{.49\linewidth}
 \begin{center}
  \includegraphics[width=.9\linewidth]{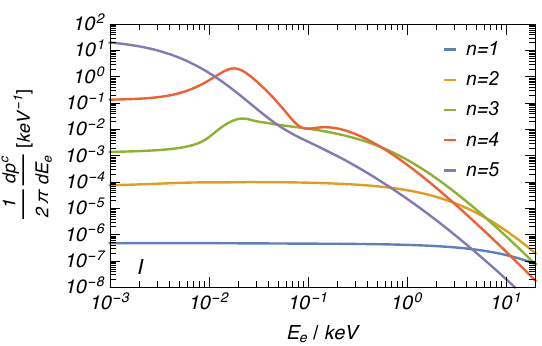}
   \end{center}
  \end{minipage}
    \caption{\sl \small
    The differential ionization probabilities as a function of the emitted electron energy, $E_e$,
for isolated Ar, Xe, Ge, Na, and I.
The contributions from different $\ell$'s are summed.
We also summed all the possible final states for a given $n$.
The integrated probabilities are given in Tab.\,\ref{tbl:excitation}.
The ionization probabilities are available in ancillary files to the arXiv version (See note added before Appendix\,\ref{sec:norm}.).
The files also contain the data of C, F,  and Ne.
}
\label{fig:dpdE}
\end{figure}
We show the numerical results of the transition probabilities to the order of $O(q_e^2)$.
The ground state of the atom consists of a complex of the orbitals given in Tab\,\ref{tbl:ground_cfg}.
If there are more than one energy eigenstate configurations 
for a given complex, we take an average of the transition probabilities
for those energy eigenstate configurations. 
As discussed in section \ref{sec:SEE}, we  consider the final states in which only one of $o_k$'s in the initial state
is replaced by an excited or an unbouded electron state.

In Fig.\,\ref{fig:dpdE}, we show the differential ionization probabilities, $dp^c_{q_e}/dE_e$,  for $q_e = m_e v_F$ with 
$v_F = 10^{-3}$.
In the figure, we sum all the contributions from different orbital angular momenta in the initial state, $\ell$, 
for a given principal quantum number, $n$.
We also sum all the possible final states for a given $n$.
It should be noted that the spectrum shape of $dp^c_{q_e}/dE_e$ does not depend on $v_F$, 
and hence, the probabilities for a different $q_e$ is obtained by multiplying $q_e^2/(m_e \times 10^{-3})^2$.
The integrated probabilities are also given in Tab.\,\ref{tbl:excitation} for a given initial $(n,\ell)$.
The results show that the ionization probabilities from the inner shells can be of ${\cal O}(10^{-2})$ for $v_F = 10^{-3}$.
The ionization probabilities from the valence electrons can be even of ${\cal O}(10^{-1})$.

{
As we have seen in Fig.\,\ref{fig:kin}, the recoil energy $E_R$ and the electron excitation energy ${\mit\D}E$ (and hence $E_e$) are 
kinematically constrained for given $m_{DM}$ and $v_{DM}$.
Cross correlations of the ionization probabilities between $E_R$ and $E_e$ are obtained by rescaling the results in Fig.\,\ref{fig:dpdE}
by $q_e^2 = 2 m_e^2E_R/m_A$ within the kinematically allowed region in Fig.\,\ref{fig:kin}.
}

In Tab.\,\ref{tbl:excitation}, we also show the excitation probabilities, $p_{q_e}^d(n\ell\to n'\ell')$.
As the table shows, the excitation probabilities are much smaller than the ionization probabilities for a given initial $n$.

{
Before closing this section, let us comment on the probability conservation in the single electron transition.
As discussed in the appendix~\ref{sec:probability}, the single electron transition probability satisfies,
\begin{eqnarray}
\frac{1}{\omega_{n,\ell}}\!\!
\left(\sum_{n',\ell'}
  p_{q_e}^d\!(n\ell\to n'\ell')\!+\!\!
\!   \int\!\!\frac{dE_e}{2\pi}\frac{d}{dE_e}p_{q_e}^c(n\ell\to E_e)\!\!\right)\!\!=\!
 1\!- p_{\rm unchanged}(n\ell)\! -\! p_{\rm occupied}(n\ell)\ .
 \label{eq:prob}
\end{eqnarray}
Here, $p_{\rm unchanged}$ and $p_{\rm occupied}$ are given in Eqs.\,(\ref{eq:unchanged}) and (\ref{eq:occupied}),
while the lefthand side corresponds to $p_{\rm ex}(n\ell)$ in Eq.\,(\ref{eq:ex}) 
(see also Eqs.\,(\ref{eq:excitationELL}) and \,(\ref{eq:ionizationELL})).
In our analysis, we numerically checked that $p_{\rm unchanged}$, $p_{\rm occupied}$, and $p_{\rm ex}$ 
satisfy the probability conservation in Eq.\,(\ref{eq:prob}).%
\footnote{Numerically, 
$p_{\rm occupied}$ is at the same order or even an order of  magnitude larger than $p_{\rm ex}$.
}

In Ref.\,\cite{Bernabei:2007jz}, the probability $p_{\rm occupied}$ ($P^1_{exc,i}$ in Ref.\,\cite{Bernabei:2007jz})
is incorrectly defined, which is related to $p_{\rm occupied}$ in this paper by
\begin{eqnarray}
\left.p_{\rm occupied}\right|_{\tiny [15]} = p_{\rm occupied}\times (1 - p_{\rm unchanged})\ .
\end{eqnarray}
Since $p_{\rm unchanged} \simeq 1$, $\left.p_{\rm occupied}\right|_{\tiny [15]}$ is much smaller than $p_{\rm occupied}$.
The ionization probability in  Ref.\,\cite{Bernabei:2007jz} is, on the other hand, estimated by 
using the (incorrect) probability conservation,%
\footnote{The one electron transition probabilities in this paper correspond to
$p_{\rm unchanged} = P^1_{ii}$, $p_{\rm occupied} = P^1_{exc,i}$, 
and $P^1_{bound,i} = \sum_{n',\ell'} p_{q_e}^d(n\ell\to n'\ell')/{\omega_{n,\ell}}$. }
\begin{eqnarray}
\left. p_{\rm ionization}\right|_{\tiny [15]}&=& 1- p_{\rm unchanged} -\left. p_{\rm occupied}\right|_{\tiny [15]}
-\frac{1}{\omega_{n,\ell}} \sum_{n',\ell'} p_{q_e}^d(n\ell\to n'\ell')\ .
 \label{eq:ionization15}
\end{eqnarray}
Since we find $p_{\rm occupied} \gtrsim p_{\rm ex}$ numerically, it shows that the ionization rates in Ref.\,\cite{Bernabei:2007jz}
are overestimated.
}

\newpage
\begin{table}[t]
\caption{\sl \small The excitation probabilities for a given initial state  $(n,\ell)$.
$\prob_{\to n'\ell'}$ is defined by $\prob_{\to n'\ell'} \equiv p_{q_e}^d (n\ell \to n'\ell')$.  
The probabilities not shown in this table are forbidden or negligibly small.
The integrated ionization probabilities are also shown in the rightmost column.
The averaged binding energies of the initial and the final orbitals are also shown which are obtained by {\tt FAC}.
}
\begin{center}
\begin{tcolorbox}[title=Ar $(q_e\defeq  m_e\times10^{-3})$,colback=white,width=1\linewidth]\footnotesize{
\begin{minipage}{1\linewidth}
\label{tbl:excitation}

\fontsize{10pt}{11pt}\selectfont
\begin{tabular}{|c||c|c|c|c|c|c||c||c|}
\hline
$(n,\ell)$&  $\prob_{\to 3d}$  & $\prob_{\to 4s}$ & $\prob_{\to 4p}$& $\prob_{\to 4d}$ & $\prob_{\to 5s}$ & $\prob_{\to 5p}$  & ${E_{n\ell} }$ [eV]  & ${\frac{1}{2 \pi} \int d E_{e} \frac{d p^{c}}{d E_{e}}}$  \\ \hline 
$1s$ & --  &  -- & $1.3 \times 10^{-7}$ & -- &--& $4.3 \times 10^{-8}$ &$ 3.2\times10^{3}$  & $ 7.3\times 10^{-5}$  \\ \hline
$2s$ & --&-- & $ 5.3\times 10^{-6}$  &-- &--& $ 1.8 \times 10^{-6}$ &$3.0 \times10^{2}$& $ 5.3 \times 10^{-4}$  \\ 
$2p$ & $4.3 \times 10^{-6}$  & $5.0 \times 10^{-6}$ &--  & $ 3.0\times 10^{-6}$ & $1.3 \times 10^{-6}$ &--&$ 2.4\times10^{2}$& $4.6 \times 10^{-3}$  \\ \hline 
$3s$  &--  &--  & $5.3 \times 10^{-7}$  & --&-- & $ 1.1 \times 10^{-6}$  & $2.7\times10$  & $ 1.4 \times 10^{-3}$\\ 
$3p$  & $7.9\times10^{-3}$   & $8.5\times10^{-3}$   &--  & $ 4.0\times 10^{-3}$ & $1.2\times10^{-3}$ & --&$1.3\times10$ &$6.4\times10^{-2}$\\ \hline
\end{tabular}

\vspace{.3cm}

\begin{tabular}{|c||C|C|C|C|C|C|}
\hline
$(n,\ell)$&$3d$ & $4s$ & $4p$ & $4d$ & $5s$ & $5p$ \\
\hline
${E_{n\ell}}$[eV]& 1.6 & 3.7  &2.5 &$0.88$&1.6& 1.2\\
\hline
\end{tabular}
\end{minipage}
}
\end{tcolorbox}
\end{center}

\vspace{-.7cm}

\begin{center}
\end{center}
\end{table}

\newpage
\setcounter{table}{1}
\begin{table}[h]
\begin{center}
\begin{tcolorbox}[title=Xe $(q_e\defeq  m_e\times10^{-3})$,colback=white,width=.8\linewidth]\footnotesize{
\begin{minipage}{1\linewidth}
\fontsize{10pt}{11pt}\selectfont
\begin{tabular}{|c||c|c|c|c||c||c|}\hline
($n,\ell$) &  $\prob_{\to 4f}$  &  $\prob_{\to 5d}$  & $\prob_{\to 6s}$  & $\prob_{\to 6p}$   & ${E_{n\ell}} $ [eV] 
 & $\frac{1}{2 \pi} \int d E_{e} \frac{d p^{c}}{d E_{e}}$
 \\ \hline 
1s & --& --  & --  & $7.3 \times 10^{-10}$& $3.5 \times 10^{4}$    & $4.9 \times 10^{-6}$ 
\\ \hline
2s &-- & --  & --  & $ 1.8\times 10^{-8}$  & $ 5.4 \times 10^{3}$    & $3.0 \times 10^{-5}$
\\ 
2p & --& $3.0 \times 10^{-8}$  & $6.5 \times 10^{-9}$ &-- & $4.9 \times 10^{3}$    & $1.3 \times 10^{-4}$ 
\\ \hline 
3s  &-- & --  & --  & $2.7 \times 10^{-7}$  & $ 1.1 \times 10^{3}$   &$1.1 \times 10^{-4}$
\\ 
3p & --& $3.4 \times 10^{-7}$   & $ 4.0\times 10^{-7}$   &--& $9.3 \times 10^{2}$   & $6.0 \times 10^{-4}$ 
\\ 
3d & $2.3 \times 10^{-9}$ & --   &--   & $4.3 \times 10^{-7}$ & $6.6 \times 10^{2}$  & $3.6 \times 10^{-3}$ 
\\  \hline
4s  & --& --  &--  &$3.1 \times 10^{-6}$   & $ 2.0 \times 10^{2}$   & $3.6 \times 10^{-4}$
\\ 
4p &--  & $4.1 \times 10^{-8}$   & $3.0 \times 10^{-5}$   & --  &$1.4 \times 10^{2}$  & $1.5 \times 10^{-3}$ 
\\ 
4d &   $ 7.0\times 10^{-7}$& -- &  -- & $1.5 \times 10^{-4}$ & $6.1 \times 10$  & $3.6 \times 10^{-2}$  
\\\hline
5s &-- & --   &-- & $1.2 \times 10^{-4}$   &  $ 2.1 \times 10$  & $4.7 \times 10^{-4}$ 
\\ 
5p & --& $3.6 \times 10^{-2}$ &   $2.1 \times 10^{-2}$    & -- &  $9.8$ & $7.8 \times 10^{-2}$ 
\\ \hline
\end{tabular}

\vspace{.2cm}

\begin{tabular}{|c||C|C|C|C|}
\hline
$(n,\ell)$&$4f$ & $5d$ & $6s$ & $6p$ \\
\hline
${E_{n\ell}}$[eV]& 0.85 & $1.6$  &$3.3$ &$2.2$\\
\hline
\end{tabular}
\end{minipage}
}
\end{tcolorbox}
\end{center}
\end{table}

\begin{table}[h]
\begin{center}
\begin{tcolorbox}[title=Ge $(q_e\defeq  m_e\times10^{-3})$,colback=white,width=0.9\linewidth]\footnotesize{
\begin{minipage}{1\linewidth}
\begin{tabular}{|c||c|c|c|c|c||c||c|}\hline
($n,\ell$) & $\prob_{\to 4p}$& $\prob_{\to 4d}$  & $\prob_{\to 5s}$  & $\prob_{\to 5p}$  & $\prob_{\to 6s}$ 
& ${E_{n\ell}}$ [eV]  & $\frac{1}{2 \pi} \int d E_{e} \frac{d p^{c}}{d E_{e}}$ 
\\ \hline 
1s &  $5.0 \times 10^{-8}$  &-- &-- & $7.9 \times 10^{-9}$ &--&$1.1\times10^{4}$ &$1.8\times10^{-5}$ 
\\ \hline
2s &  $1.8 \times 10^{-6}$  &-- &--  & $2.8 \times 10^{-7}$ &--&$1.4\times10^{3}$ &$1.3\times10^{-4}$ 
\\ 
2p &-- & $3.3 \times 10^{-7}$ &$1.1 \times 10^{-7}$    &--& $3.4 \times 10^{-8}$&$1.2\times10^{3}$&$7.3\times10^{-4}$ 
 \\ \hline 
3s  & $3.7 \times 10^{-5}$   &--  &--  &$5.6 \times 10^{-6}$ & -- &$1.7\times10^{2}$&$5.5\times10^{-4}$ 
\\ 
3p  &--  & $ 6.0\times 10^{-9}$   &$2.8 \times 10^{-5}$    &--& $8.3 \times 10^{-6}$  & $1.2\times10^{2}$
&$2.4\times10^{-3}$
\\ 
3d   & $ 2.3 \times 10^{-3}$   &--  &-- & $2.3 \times 10^{-4}$ & -- & $3.5\times10$&$2.8\times10^{-2}$
\\  \hline
4s   & $4.0 \times 10^{-2}$  &-- &--    & $3.9 \times 10^{-4}$ & --& $1.5\times10$&$6.1\times10^{-4}$ 
\\ 
4p   & -   & $2.7 \times 10^{-2}$   & $1.6 \times 10^{-2}$   & -- &$1.5 \times 10^{-3}$&6.5&$2.6\times10^{-2}$ 
 \\ 
\hline
\end{tabular}

\vspace{.2cm}

\begin{tabular}{|c||C|C|C|C|}
\hline
$(n,\ell)$& $4d$ & $5s$ & $5p$ & $6s$ \\
\hline
${E_{n\ell}}$[eV]& $1.6$ & $3.0$  &$2.0$ &$1.4$\\
\hline
\end{tabular}
\end{minipage}
}
\end{tcolorbox}
\end{center}

\end{table}

\setcounter{table}{1}
\begin{table}[h]
\begin{center}
\begin{tcolorbox}[title=Na $(q_e\defeq  m_e\times10^{-3})$,colback=white,width=1.0\linewidth]\footnotesize{
\begin{minipage}{1\linewidth}
\begin{tabular}{|c||c|c|c|c|c|c||c||c|}\hline
($n,\ell$)&  $\prob_{\to 3s}$  & $\prob_{\to 3p}$ & $\prob_{\to 3d}$ & $\prob_{\to 4s}$ & $\prob_{\to 4p}$ & $\prob_{\to 4d}$ 
 & ${E_{n\ell}}$ [eV] & $\frac{1}{2 \pi} \int d E'_{k} \frac{d p^{c}}{d E'_{k}}$
 \\ \hline 
1s & -- & $2.1 \times 10^{-6}$  &-- &--& $6.4 \times 10^{-7}$ & --&$1.1\times10^{3}$ &$2.5 \times10^{-4}$ 
\\ \hline
2s & -- & $6.8 \times 10^{-5}$  &--  &--& $ 2.0 \times 10^{-5}$ & --&$6.5\times10$&$1.7 \times10^{-3}$ 
\\ 
2p & $5.9 \times 10^{-5}$  &--&$1.1 \times 10^{-4}$    & $1.5 \times 10^{-4}$ &--&$6.2 \times 10^{-5}$ 
&$3.8\times10$
&$2.2 \times10^{-2}$ 
\\ \hline 
3s  & -   & $8.8 \times 10^{-2}$   &--  & --& $ 1.1 \times 10^{-3}$ &--&$6.1$& $5.3\times10^{-4}$
 \\ \hline
\end{tabular}

\vspace{.2cm}

\begin{tabular}{|c||C|C|C|C|C|}
\hline
$(n,\ell)$&$3p$ & $3d$ & $4s$ & $4p$ & $4d$  \\
\hline
${E_{n\ell}}$[eV]& 3.3 & 1.5  &2.1 &$1.5$&0.86\\
\hline
\end{tabular}

\end{minipage}
}
\end{tcolorbox}

\end{center}
\end{table}

\begin{table}
\begin{center}
\begin{tcolorbox}[title=I $(q_e\defeq  m_e\times10^{-3})$,colback=white,width=0.9\linewidth]\footnotesize{
\begin{minipage}{1\linewidth}
\begin{tabular}{|c||c|c|c|c|c||c||c|}
\hline
($n,\ell$) &$\prob_{\to 4f}$& $\prob_{\to 5p}$& $\prob_{\to 5d}$  & $\prob_{\to 6s}$  & $\prob_{\to 6p}$   
 & ${E_{n\ell}} $ [eV]  & $\frac{1}{2 \pi} \int d E_{e} \frac{d p^{c}}{d E_{e}}$
\\ \hline 
1s &-- & $2.0 \times 10^{-9}$  &-- &-- & $7.8 \times 10^{-10}$ &$3.3\times10^{4}$ & $5.1 \times 10^{-6}$ 
\\ \hline
2s &-- & $5.0 \times 10^{-8}$  &-- &--  & $2.0 \times 10^{-8}$  &$5.1\times10^{3}$ & $ 3.1 \times 10^{-5}$
 \\ 
2p &--&-- & $3.3 \times 10^{-8}$ &$6.9 \times 10^{-9}$    &--&$4.6\times10^{3}$ & $1.4 \times 10^{-4}$
\\ \hline 
3s  &--& $7.7 \times 10^{-7}$   &--  &--  &$3.0 \times 10^{-7}$&$1.0\times10^{3}$ & $ 1.2 \times 10^{-4}$
 \\ 
3p & --&--  & $ 3.8\times 10^{-7}$   &$4.4 \times 10^{-7}$    &-- &$8.7\times10^{2}$& $6.4 \times 10^{-4}$
\\ 
3d &  $ 1.7 \times 10^{-9}$ & $ 1.3 \times 10^{-6}$   &--  &-- & $5.0 \times 10^{-7}$ &$6.1\times10^{2}$ 
& $3.8 \times 10^{-3}$
\\  \hline
4s  &-- & $9.2 \times 10^{-6}$  &--  &--    & $3.4 \times 10^{-6}$ &$1.8\times10^{2}$ & $ 3.8 \times 10^{-4}$
\\ 
4p  & --&--  & $1.6 \times 10^{-7}$   & $3.6 \times 10^{-5}$   &--&$1.3\times10^{2}$&$1.5 \times 10^{-3}$
 \\ 
4d  &  $ 9.8\times 10^{-7}$& $ 7.7\times 10^{-4}$ &  --  &-- & $2.0 \times 10^{-4}$ &$5.1\times10$ 
 & $4.0 \times 10^{-2}$
\\
5s &--&$8.9 \times 10^{-3}$   &--&--   & $1.8 \times 10^{-4}$&$1.9\times 10$ &  $ 4.7 \times 10^{-4}$ 
\\ 
5p &-- & - &   $4.0 \times 10^{-2}$    & $2.2 \times 10^{-2}$ &--&8.8&  $6.6\times10^{-2}$ 
\\ 
\hline
\end{tabular}

\vspace{.2cm}

\begin{tabular}{|c||C|C|C|C|}
\hline
$(n,\ell)$& $4f$ & $5d$ & $6s$ & $6p$ \\
\hline
${E_{n\ell}}$[eV]& $0.85$ & $1.6$  &$3.2$ &$2.1$\\
\hline
\end{tabular}
\end{minipage}

}
\end{tcolorbox}

\end{center}
\end{table}

\clearpage

\section{Effects on  dark matter direct detection}
\label{sec:DM}
As we have shown in the previous sections, a nuclear recoil is accompanied by the ionization and the excitation of the atom through the Migdal effect.
The electric energy released by the ionization and the de-excitation can be in the keV range when the incident dark matter velocity exceeds the threshold in Eq.\,\eqref{eq:vDMth}.

In this section, we discuss how those electronic energy injections 
affect the dark matter signals at direct detection experiments. 
In our analysis, we have assumed isolated atoms.
Thus, the results in the previous section are not directly applicable to the non-isolated atoms in liquid or crystals.
For example, the energy levels of the valence electrons are affected by the ambient atoms by ${\cal O}(0.1)$\,eV.%
\footnote{For the energy levels of the valence electrons of the liquid Xe, see e.g. Ref.~\cite{Mulliken1970}. }
Furthermore, when the electronic band structure is formed in the medium, the excitations into the unoccupied state
should be reinterpreted as transitions into the conducting band.

The ionization rates from the inner orbitals are, on the other hand, expected to be less affected by 
the ambient atoms.
In fact, the binding energies of the inner orbital are much larger than eV,
and hence, the relevant length scales for the transition factors, $z_e$, are much smaller than the typical distance between atoms.
Thus, the ionization spectrum in Eq.\,(\ref{eq:ionization}) can be applied rather reliably for the ones from the inner orbitals.%
\footnote{ As it is highly difficult to quantify the uncertainties from the effects of the ambient atoms,
it is desirable to test the Migdal effect experimentally via the low energy nuclear recoils with $E_R \ll {\cal O}(1)$\,keV.
}
In the following, we confine our arguments to  liquid Xe detectors.

\subsection{Migdal Effects on the Recoil Spectrum}
In the absence of the Migdal effect, the liquid Xe detectors respond 
to the nuclear recoil roughly through the following steps~\cite{PhysRevB.46.11463,Aprile:2009dv,2009physics}.
After the nuclear recoil, the electron clouds are assumed to catch up with the nucleus immediately,
so that the atom remains neutral.
The recoil ``atom'' loses its energy through scattering with adjacent atoms 
in the medium where the inelastic scatterings involve the ionization and excitation of the atoms.
These processes continue until the scattered atoms are thermalized.
Eventually, a fraction of the initial recoil energy  $E_R$ is converted to the measurable electronic excitation
while the rest is lost into the heat of the medium~\cite{Lindhard,Hitachi:2005ti,Sorensen:2011bd,Sorensen:2014sla}.

This should be contrasted with the electron recoils caused by  incident gamma or beta rays,
where the entire recoil energy is transferred to the measurable electronic excitation. 
Conventionally, the calibrated nuclear recoil energy, $E_{nr}$, and the electron equivalent energy, $E_{ee}$,
measured by the scintillation photons are related by
\footnote{See e.g. Ref.\,\cite{Dahl:2009nta} for the details of  energy scale calibration of the liquid Xe detectors.}
\begin{eqnarray}
\label{eq:Leff}
E_{nr} = \frac{E_{ee}}{{\cal L}_{\rm eff}} \cdot \frac{S_e}{S_n}\ .
\end{eqnarray}
Here, ${\cal L}_{\rm eff}$ is defined 
as the ratio between $E_{ee}$ and $E_{nr}$ at zero drift field relative to $122$\,keV gamma rays,
which is ${\cal L}_{\rm eff} \simeq 0.1-0.2$ for 
$E_{nr} \lesssim 100$\,keV~\cite{Sorensen:2010hv,Horn:2011wz,Aprile:2013teh,Akerib:2016mzi}.
The quantities, $S_e$ and $S_n$, are the scintillation quenching factors  of electron and nuclear recoils
due to the drift electric field, $E_d$, which are $S_e\simeq 0.4$--$1$~\cite{Aprile:2005mt} and $S_n \simeq 1$~\cite{Manzur:2009hp}
for $E_d \le 4$\,kV/cm, respectively.%
\footnote{For $E_d = 0$\,kV/cm, {$S_e= S_n = 1$.}}

In the presence of the transition due to the Migdal effect, the nuclear recoil is accompanied by  electronic energy injections in the 
sub-keV to the keV range.
By ignoring the energy resolution of the detectors, the electron equivalent energy spectrum 
is  given by
\begin{eqnarray}
\label{eq:single}
\frac{dR}{dE_{\rm det}dv_{DM}} &\simeq&
\int dE_R dE_{EM} \,\delta(E_{\rm det} - q_{nr}E_R - E_{EM})
\frac{dR}{dE_{R}\, dE_{EM}\, dv_{DM}}\ ,
\end{eqnarray}
with $q_{nr}$ being the conversion between $E_{nr}$ and $E_{ee}$ in Eq.\,\eqref{eq:Leff}
(see also Eqs.\,\eqref{eq:EEM} and \eqref{eq:deex2}).
In the following, we include only the electronic energy injection caused by the ionizations
as the excitation probabilities into the unoccupied binding energy levels are much smaller than the ionization probabilities.%
\footnote{Moreover, the excitation into the unoccupied binding energy levels are not well defined 
when the electronic band structure is formed in the medium.
}

\begin{figure}[tb]
\begin{minipage}{.49\linewidth}
\begin{center}
  \includegraphics[width=.9\linewidth]{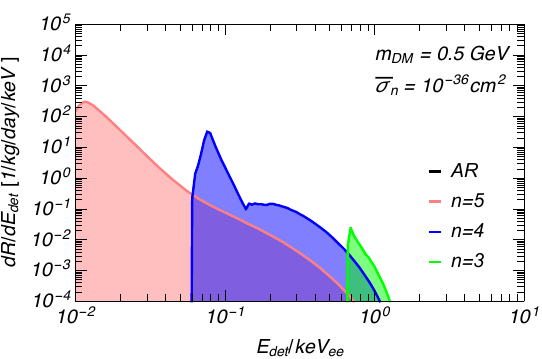}
  \end{center}
  \end{minipage}
 \begin{minipage}{.49\linewidth}
 \begin{center}
  \includegraphics[width=.9\linewidth]{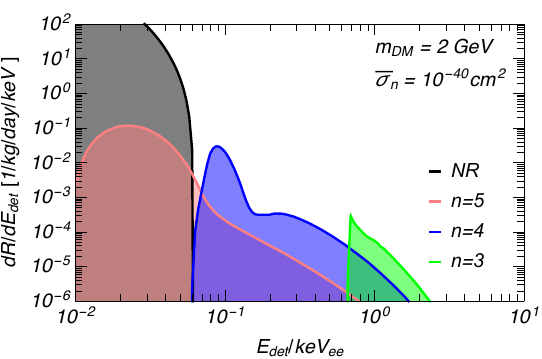}
   \end{center}
  \end{minipage}
  \begin{minipage}{.49\linewidth}
\begin{center}
  \includegraphics[width=.9\linewidth]{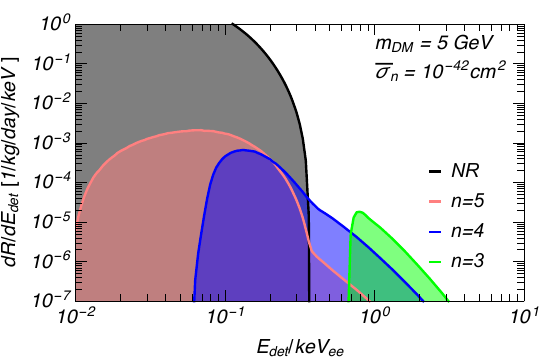}
  \end{center}
  \end{minipage}
 \begin{minipage}{.49\linewidth}
 \begin{center}
  \includegraphics[width=.9\linewidth]{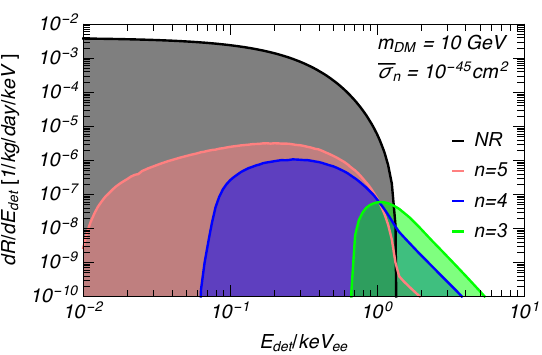}
   \end{center}
    \end{minipage}
      \begin{minipage}{.49\linewidth}
\begin{center}
  \includegraphics[width=.9\linewidth]{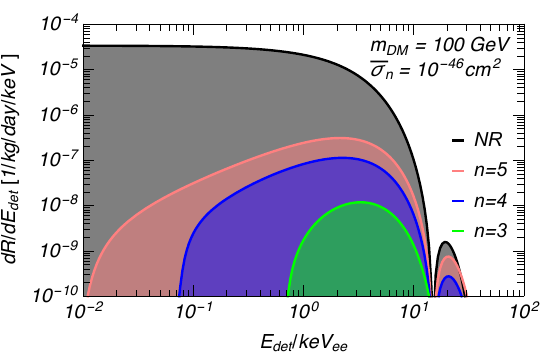}
  \end{center}
  \end{minipage}
 \begin{minipage}{.49\linewidth}
 \begin{center}
  \includegraphics[width=.9\linewidth]{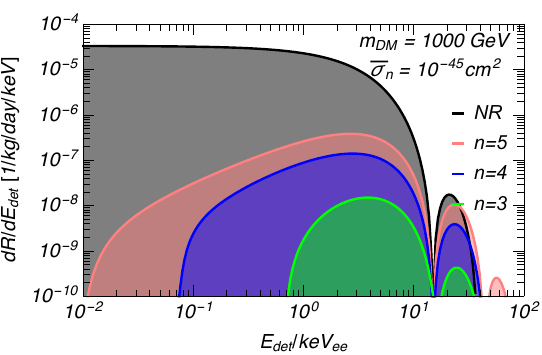}
   \end{center}
    \end{minipage}
\caption{\sl \small The differential event rates expected at the single-phase experiments with the liquid Xe target.
The black lines show the nuclear recoil (NR) spectrum without ionization.
(The NR spectrum with respect to $E_{\rm det}$ 
and the one with respect to $E_R$  differ by  a factor of $1/q_{nr}$.)
The green, blue, and pink lines show the ionization rates from $n = 3, 4$, and $5$, respectively.
Here, we do not take the energy resolution into account.
Since we apply the estimations for the isolated atoms, the ionization spectrum from 
the valence electrons, i.e. $n = 5$, are not reliable. 
}
\label{fig:Single}
\end{figure}

In Fig.\,\ref{fig:Single}, we show  the total electron equivalent energy spectrum 
for a given spin-independent scattering cross section of dark matter  on nucleons
through the contact interaction.%
\footnote{The nucleon-dark matter cross section $\bar\sigma_n$ is related to $\bar\sigma_N$ via,
$\bar\sigma_N =A^2\times \mu_N^2/\mu_n^2 \times \bar\sigma_n$ { for $g_p = p_n$}.}
Here, we assume a detector with $E_d = 0\,$kV/cm as in the single-phase experiment such as the XMASS experiment~\cite{Abe:2013tc}.
We also fix $q_{nr}= {\cal L}_{\rm eff} = 0.15$.
We adopt the Helm form factor~\cite{Lewin:1995rx,Jungman:1995df}. 
The local dark matter density is fixed to a conventional value, $\rho_{DM} = 0.3$\,GeV/cm$^{3}$. 
The local circular velocity is also fixed to be $v_{\rm circ} = 220$\,km/s with the peculiar motions 
of the Earth neglected.%
\footnote{The annual modulation caused by the Earth's peculiar motion can be significantly enhanced as of the
inelastic nuclear scattering~\cite{TuckerSmith:2001hy}.}
We also assume a Maxwell velocity distribution with the velocity dispersion, $v_0 = 220$\,km/s, 
which is truncated at the Galactic escape velocity $v_{\rm esc} = 544$\,km/s.

The figures show that the electronic energy from the ionizations can be larger than 
the maximum value of the (electron equivalent)  nuclear recoil energy for a rather light dark matter.
As discussed in the previous section, the shape of the energy spectrum of the electronic injections is 
not sensitive to the incident dark matter velocity as long as they are kinematically allowed.
The nuclear recoil energy, on the other hand, depends on the dark matter velocity, 
\begin{eqnarray}
E_R \simeq \frac{q_A^2}{2\physicalMass} \lesssim \frac{1}{2}\frac{\mu_N^2}{\physicalMass} v_{DM}^2\ ,
\end{eqnarray}
which is suppressed for light dark matter.
These features can be seen from  the figures;  the electronic energy from the ionizations 
are less sensitive to the dark matter mass, 
while the maximum nuclear recoil energy is sensitive.

Given a typical threshold  of the electron equivalent energy of the liquid Xe detectors of about a keV$_{ee}$,
the ionization from $n = 3$ provides a new detection channel for rather light dark matter.
For example, with exposures of about $10^5$\,kg$\cdot$days, 
a few hundred events are expected for $m_{DM} \simeq 500$\,MeV and $\bar\sigma_n \simeq 10^{-36}$cm$^2$
in the liquid Xe detectors.%
\footnote{For much lighter dark matter, the ionizations from $n_I = 3$ require very fast dark matter 
(see Fig.\,\ref{fig:phase}), and hence, the event rate is highly suppressed due to the dark matter velocity distribution.}
Similarly, a few signal events are expected   for $m_{DM} \simeq 5$\,GeV for $\bar\sigma_n \simeq 10^{-42}$cm$^2$
for the same exposure.
It should be noted that those signals are eliminated as background events in the conventional analysis of the dual-phase 
experiments. 
In those experiments, one needs different analyses to cover such signals (see e.g.~\cite{Essig:2012yx,Essig:2017kqs}).

For heavier dark matter, $m_{DM} >{\cal O}(10)$\,GeV, on the other hand, the Migdal effect is submerged below 
the conventional nuclear recoil spectrum, and hence, does not affect the detector sensitivities.
In principle, the additional electronic energy injections affect the so-called S2/S1 ratio in the dual-phase detectors.
This is because the numbers of direct excitons $N_{ex}$ and direct ionizations $N_i$ are different 
for the atomic recoil and the electronic recoil.%
\footnote{For the electronic recoil, the ratio is given by $N_{ex}/N_i\simeq 0.06$~\cite{PhysRevA.12.1771}, 
while it is $N_{ex}/N_i \sim 1$~\cite{2009physics,Sorensen:2011bd} for the atomic recoil.}
However, such effects should have been taken into account by the detector calibration by the neutron sources
for a given momentum transfer.

Finally, let us compare the Migdal effect with the final state photon emission of the nuclear scattering~\cite{Kouvaris:2016afs} (see also~\cite{McCabe:2017rln}).
Similarly to the Migdal effect, the final state emission  is also a universal effect and irreducible. 
The expected rates are, however, subdominant compared with the Migdal effect for $E_R$ in the keV range~\cite{Bell:2019egg}.
The Migdal effect should also be distinguished from the photon emissions in the inelastic nuclear scatterings
which require a larger momentum transfer~\cite{Ellis:1988nb,Bernabei:2000qn}. 

{\section{Migdal Effects in Coherent Neutrino-Nucleus Scattering}
\label{sec:neutrino}

As another application, let us briefly discuss the Migdal effects  in the coherent neutrino-nucleus scattering (C$\nu$NS).
In a similar manner to the dark matter scattering cross section, 
the differential cross section of the coherent neutrino-nucleus  scattering 
with the Migdal effect is given by
\begin{eqnarray}
\frac{d\sigma}{dE_R } 
 &\simeq& 
 \sum_{E_{ec}^F} 
\frac{d\sigma_{C\nu NS}}{dE_R } 
\times 
 |Z_{FI}(q_e)|^2\ . 
 \end{eqnarray}
Here, $\sigma_{C\nu NS}$ denotes the coherent neutrino-nucleus scattering~\cite{Freedman:1973yd,Freedman:1977xn,Drukier:1983gj}, 
 \begin{eqnarray}
 \frac{d\sigma_{C\nu NS}}{dE_R } & = & 
  \frac{
|F_A(q_A^2)|^2 Q_W^2  G_F^2 m_A}{4\pi}
\left(
1 
- \frac{ m_A E_R }{2 E_\nu^2}
\right)\ ,
\end{eqnarray}
with $G_F$, $\sin\theta_W$, $N$, $Q_W$ being the Fermi constant, the weak mixing angle, the number of neutrons $N=A-Z$, and the weak charge of the nucleus
$Q_W = N- (1-4 \sin^2\theta_W) Z\simeq N$, respectively.
As in the previous section, the magnetic quantum numbers of the electrons in the initial/final states are averaged/summed.
For given $E_\nu$ and ${\mit\Delta}E$, the recoil energy is constrained in
\begin{eqnarray}
\frac{{\mit \Delta}E^2}{2m_A}< E_R < \frac{(2E_\nu - {\mit\Delta}E)^2 }{2(m_A +2 E_\nu)}\ ,
\end{eqnarray}
where the maximal recoil energy corresponds to the back-to-back scattering.

In Fig.\,\ref{fig:solarXe}, we show the total electron equivalent energy spectra for the C$\nu$NS
for the pp and the $^8$B solar neutrinos.
Here, we take the central values of the neutrino fluxes of the SFII-GS98 model given in Ref.\,\cite{Serenelli:2016dgz}.
The single-phase liquid Xe detectors are assumed as in the previous section with $q_{nr} = 0.15$.
The figures show that the nuclear recoil signal without the Migdal effect is below the energy thresholds of 
the current liquid Xe detectors, i.e. ${\cal O}(1)$\,keV.
The signal of the ionization from $n_I = 3$ is, on the other hand, above the energy threshold,
a few events of which are expected with exposures of about $10$\,ton$\cdot$years.%
\footnote{More detailed study including background estimation will be given elsewhere.}

\begin{figure}[tb]
\begin{minipage}{.49\linewidth}
\begin{center}
  \includegraphics[width=.9\linewidth]{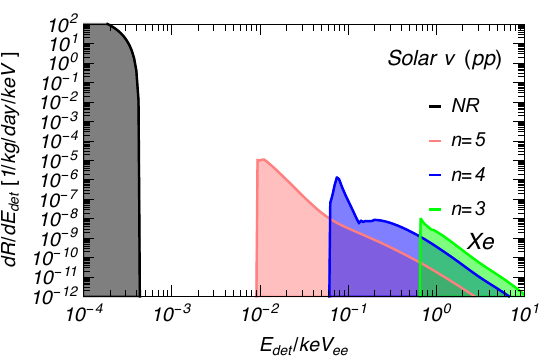}
  \end{center}
  \end{minipage}
  \begin{minipage}{.49\linewidth}
\begin{center}
  \includegraphics[width=.9\linewidth]{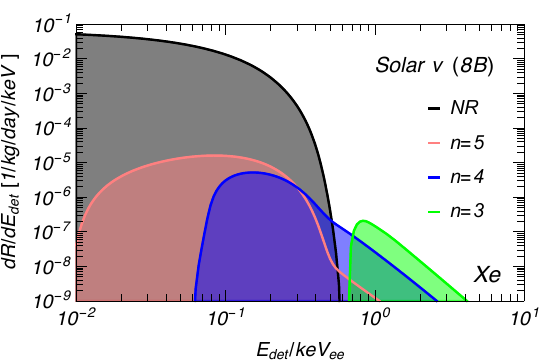}
  \end{center}
  \end{minipage}
\caption{\sl \small 
The differential event rates for expected  for the coherent neutrino-nucleus scattering for the pp and the $^8$B solar neutrinos. 
The single-phase liquid Xe detectors are assumed as in the previous section with $q_{nr} = 0.15$.
The black lines show the nuclear recoil (NR) spectrum without the ionizations.
(The NR spectrum with respect to $E_{\rm det}$ 
and the one with respect to $E_R$  differ by  a factor of $1/q_{nr}$.)
Since we apply the estimations for the isolated atoms, the ionization spectrum from the valence electrons are not reliable. 
}
\label{fig:solarXe}
\end{figure}
}

\section{Conclusions and Discussion}
In this paper, we reformulated the Migdal effect at the nuclear recoil caused by a dark matter scattering
and a coherent neutrino-nuclear scattering.
In our formalism, we take the plane waves of the whole atomic system as the asymptotic states 
for the scattering process.
The coherent treatment of the electron cloud makes the energy-momentum conservation and the probability conservation transparent. 
We also provide numerical estimates of the ionization and the excitation rates 
for isolated atoms of Ar, Xe, Ge, Na, and I by using the {\tt Flexible Atomic Code (FAC, cFAC)}~\cite{Gu2008}.

We also applied the results for the dark matter direct detections by taking a liquid Xe detector as an example.
We showed that the ionization signals through the Migdal effect provide new detection channels for 
light dark matter with a mass in the GeV range.
Since such signals are eliminated as background events in the conventional analysis of the dual-phase experiments, 
different analyses are required to cover such signals.
For rather heavy dark matter, on the other hand, the Migdal effects are submerged below the 
conventional atomic recoil spectrum.

In our analysis, we have not studied detailed detector responses nor the precise treatment 
of the Migdal effects of the non-isolated atoms in liquid or crystals.
For more precise estimation, detailed detector simulations are imperative.
In particular, it is important to study detector responses to the 
energy released by the de-excitation of the core-hole. 
More theoretical efforts are also required for  precise estimation of the Migdal effect 
in the medium.

We also note that it is an interesting future work to discuss whether the Migdal effect 
affects the direct detection experiments that mainly aim the electron recoil~\cite{Essig:2011nj,
Graham:2012su,Hochberg:2015pha,Lee:2015qva,Essig:2015cda,Hochberg:2015fth,Hochberg:2016ntt,Kadribasic:2017obi,Cavoto:2017otc}.
As the Migdal effect provides the electronic signals via the nuclear recoil, 
dark matter without electron recoil can be searched for by those experiments.

\section*{Acknowledgements}
\vspace{-.3cm}
The authors acknowledge Y.~Itow, Y.~Kishimoto, and S.~Moriyama for useful discussion.
The authors also acknowledge H.~Ejiri for his informative seminar at ICRR, which drew our attention to the Migdal effect.
This work is supported in part by Grants-in-Aid for Scientific Research from the Ministry of Education, Culture, Sports, Science, and Technology (MEXT) KAKENHI, Japan, No.\,25105011, No.\,15H05889 and  No.\,17H02878 (M. I.).

\section*{Note Added}
The ionization probabilities used in Fig.\,\ref{fig:dpdE} are available in  ancillary files to the arXiv version.
The probabilities are obtained by using the {\tt Flexible Atomic Code (FAC, cFAC)}~\cite{Gu2008}.
The files contain free electron energy in the unit of eV and the differential  probability of $dp^c/dE_e$ in the unit of eV$^{-1}$
for a given ($n,\ell$).
Note that we do not divede the data by $2\pi$.
The probabilities need to be rescaled by $(q_e/{1\,{\rm eV}})^2$ for $q_e \neq 1$\,eV.

\appendix
\section{The Normalization of the Projection Operator}
\label{sec:norm}
In this appendix, we show that the total projection operator is given by
\begin{eqnarray}
\label{eq:projection0}
\int d\hat P =
\int\frac{d^3{\mathbf p}_{A}}{(2\pi)^3}
\sum_{E_{ec}^F} 
|\Psi_E\rangle\langle \Psi_E| = \mathbbm{1} \ ,
\end{eqnarray}
where the summation is taken for all the possible electron cloud configurations including the continuous spectrum.
In terms of the one-electron states, the projection operators can also be written by
\begin{eqnarray}
\label{eq:projection}
\int d\hat P = 
\int\frac{d^3{\mathbf p}_{A}}{(2\pi)^3 }
\left(
|{\mathbf p}_{N}\rangle
\langle{\mathbf p}_{N}|
\right)
\left(\prod_{i=1}^{N_e}\,\,\,\mathclap{\displaystyle\int}\mathclap{\textstyle\sum}\,\,\,_{o_i}
|\tilde\phi_{o_i}\rangle \langle\tilde \phi_{o_i} |
\right)
= \mathbbm{1} \ ,
\end{eqnarray}
where 
\begin{eqnarray}
|{\mathbf p}_{N}\rangle &= & 
\int d^3 \XN \, |\XN\rangle\, e^{i{\mathbf p}_N \cdot \XN}
\ ,\\
|\tilde \phi_{o_i} \rangle &=& 
e^{i{\mathbf q_e}\cdot \hat{\mathbf{x}}_i - i \hXN\cdot\hat{\mathbf p}_i}|\phi_{o_i}\rangle \ , \\
\label{eq:pNpA}
{\mathbf p}_{N} & = & m_N {\mathbf v} = \frac{m_N}{\nominalMass} {\mathbf p}_{A}\ ,\\
\label{eq:qepA}
{\mathbf q}_e &= & m_e  {\mathbf v} = \frac{m_e}{\nominalMass} {\mathbf p}_{A}\ , \\
\,\,\,\mathclap{\displaystyle\int}\mathclap{\textstyle\sum}\,\,\,_{o_i} & = & \sum_{n_i,\kappa_i,m_i } + \sum_{\kappa_i,m_i} \int \frac{dE_i}{2\pi} \ .
\end{eqnarray}
Here, $|\phi_{o_i}\rangle$ denotes the energy eigenstate for a single orbital which corresponds to
\begin{eqnarray}
\langle \xei, \alpha_i |\phi_{o_i}\rangle = \phi^{\alpha_i}_{o_i}(\xei)\ ,
\end{eqnarray}
in the coordinate representations with the spinor index $\alpha_i$.
Eq.\,(\ref{eq:pNpA}) and Eq.\,(\ref{eq:qepA}) represent the relation between the parameters, not the operator identities.
The one particle states   are normalized such that
\begin{eqnarray}
\langle \phi_{o} | \phi_{o'}\rangle &=& \delta_{nn'}\delta_{\k\k'} \delta_{mm'}\,\,(E_o<0)\ ,\\
\langle \phi_{o} | \phi_{o'}\rangle &=& (2\pi)\delta(E_o-E_o')\delta_{\k\k'} \delta_{mm'} \,\, (E_o>0) \ ,\\
\langle{\mathbf p}_{N}|{\mathbf p'}_{N}\rangle &=&  (2\pi)\delta^3({\mathbf p}_N-{\mathbf p}_N') \ .
\end{eqnarray}

In this notation, the energy eigenstate in Eq.\,(\ref{eq:eigenPSI}) is given by
\begin{eqnarray}
\label{eq:|PSE>}
|\Psi_{E}\rangle =
\left( \sum_{\sigma \in S_{N_e}} 
 \frac{\rm sgn(\sigma)}{\sqrt{N_e!}}
\Pi_{i=1}^{N_e}e^{i{\mathbf q_e}\cdot \hat\xe_i - i \hXN\cdot\hat{\mathbf p}_i}  |\phi_{{o}_{\sigma(i)}}\rangle
\right)|{\mathbf p}_N\rangle\ .
\end{eqnarray}
Note again that this wave function is not the eigenstate of nucleus momentum $\hat{{\bf p}}_N$.
By applying the operator  in Eq.\,\eqref{eq:projection} on Eq.\,\eqref{eq:|PSE>}, we obtain,
\begin{eqnarray}
\label{eq:step1}
\int d\hat P |\Psi_E\rangle &= &
\int\frac{d^3{\mathbf p}'_{A}}{(2\pi)^3 }
|{\mathbf p}'_{N}\rangle
\nonumber\\
&&\times\langle{\mathbf p}'_{N}|
\left( \sum_{\sigma \in S_{N_e}} 
 \frac{\rm sgn(\sigma)}{\sqrt{N_e!}}
\Pi_{i=1}^{N_e}
\,\,\,\,\mathclap{\displaystyle\int}\mathclap{\textstyle\sum}\,\,\,_{o_i}
|\tilde\phi_{o_i}\rangle
\langle\tilde \phi_{o_i} |e^{i{\mathbf q_e}\cdot \hat\xe_i - i \hXN\cdot\hat{\mathbf p}_i}  |\phi_{{o}_{\sigma(i)}}\rangle
\right)|{\mathbf p}_N\rangle\ .
\end{eqnarray}
By inserting the projection operator,
\begin{eqnarray}
\mathbbm{1} = \int d^3 \XN  |\XN\rangle \langle \XN|  \times \prod_{i=1}^{N_e} d^3\xei\sum_{\a_i}
|\xei,\a_i\rangle \langle \xei,\a_i | \ ,
\end{eqnarray}
Eq.\,\eqref{eq:step1} is reduced to
\begin{eqnarray}
\label{eq:step2}
\int&&\frac{d^3{\mathbf p}'_{A}}{(2\pi)^3 }
|{\mathbf p}'_{N}\rangle 
d^3\XN 
e^{-i({\mathbf p}'_N-{\mathbf p}_N)\cdot \XN}\nonumber\\
&&\times \left( \sum_{\sigma \in S_{N_e}} 
 \frac{\rm sgn(\sigma)}{\sqrt{N_e!}}
\Pi_{i=1}^{N_e}
\,\,\,\mathclap{\displaystyle\int}\mathclap{\textstyle\sum}\,\,\,_{o_i}
|\tilde\phi_{o_i}\rangle
d^3\xei\sum_{\a_i }
e^{-i({\mathbf q_e}'-{\mathbf q_e})\cdot \xei }
\phi_{{o}_{i}}^{\a_i*}(\xei-\XN)\phi_{{o}_{\sigma(i)}}^{\a_i}(\xei-\XN) \right)
\nonumber  \\
=  \int&&\frac{d^3{\mathbf p}'_{A}}{(2\pi)^3 }
|{\mathbf p}'_{N}\rangle
d^3\XN 
e^{-i(({\mathbf p}'_N+N_e {\mathbf q_e}')-({\mathbf p}_N+N_e{\mathbf q_e}))\cdot \XN}\nonumber\\
&&\times \left( \sum_{\sigma \in S_{N_e}} 
 \frac{\rm sgn(\sigma)}{\sqrt{N_e!}}
\Pi_{i=1}^{N_e}
\,\,\,\mathclap{\displaystyle\int}\mathclap{\textstyle\sum}\,\,\,_{o_i}
|\tilde\phi_{o_i}\rangle
d^3\xei\sum_{\a_i }
e^{-i({\mathbf q_e}'-{\mathbf q_e})\cdot \xei }
\phi_{{o}_{i}}^{\a_i*}(\xei)\phi_{{o}_{\sigma(i)}}^{\a_i}(\xei) \right)\ ,
\end{eqnarray}
where we have shifted the integration variable $\xei$ by $\XN$.
By remembering 
\begin{eqnarray}
{\mathbf p}_N+N_e{\mathbf q_e} =  {\mathbf p}_{A} \ , 
\end{eqnarray}
we find 
\begin{eqnarray}
\int d^3{\mathbf p}'_{A}|{\mathbf p}'_{N}\rangle&& \delta^3({\mathbf p}_{A}' - {\mathbf p}_{A}) \nonumber\\
&&\times \left( \sum_{\sigma \in S_{N_e}} 
\frac{\rm sgn(\sigma)}{\sqrt{N_e!}}
\Pi_{i=1}^{N_e}
\,\,\,\mathclap{\displaystyle\int}\mathclap{\textstyle\sum}\,\,\,_{o_i}
|\tilde\phi_{o_i}\rangle
d^3\xei\sum_{\a_i}
\phi_{{o}_{i}}^{\a_i*}(\xei)\phi_{{o}_{\sigma(i)}}^{\a_i}(\xei) \right)\ ,
\end{eqnarray}
where we have used ${\mathbf q_e}' = {\mathbf q_e}$ for ${\mathbf p}_{A}' = {\mathbf p}_{A}$.
Finally, by using the orthogonality of the electron orbitals, we confirm that 
\begin{eqnarray}
\int d\hat P |\Psi_E \rangle  = |\Psi_E\rangle \ .
\end{eqnarray}

\section{Probability Conservation and Occupied-Occupied Transition}
\label{sec:probability}
Let us discuss the probability conservation,
\begin{eqnarray}
\sum_{E_{ec}^F} 
 |Z_{FI}({\mathbf q_e})|^2 = 1 \ ,
\end{eqnarray}
in terms of the single electron transition amplitudes. 
As discussed in sec.\,\ref{sec:SEA}, the electron excitation/ionization amplitude is reduced to
\begin{eqnarray}
\langle \Phi_{E_{ec}^F}|e^{-i\sum{\mathbf q_e}\cdot \hat \xe_i}| \Phi_{E_{ec}^I} \rangle \simeq 
-i\langle \phi_{o_k'} |  {\mathbf q_e}\cdot{ \hat \xe} |\phi_{o_k } \rangle \ ,
\end{eqnarray}
at the leading order of $q_e$.
Here we use the notation in the appendix\,\ref{sec:norm}.
Accordingly, the excitation probability is given by
\begin{eqnarray}
\prob_{\rm ex} &=& \sum_{E_{ec}^F} |\langle \Phi_{E_{ec}^F}|e^{-i\sum{\mathbf q_e}\cdot \hat \xe_i}| \Phi_{E_{ec}^I} \rangle |^2 
\simeq \sum_{k=1}^{N_e}
\,\,\,\mathclap{\displaystyle\int}\mathclap{\textstyle\sum}\,\,\,_{o\cancel{\in}ec_I}|\langle \phi_{o} | i {\mathbf q_e}\cdot{ \hat \xe} |\phi_{o_k } \rangle|^2 \ .
\end{eqnarray}
Hereafter, the operator $\hat{\bf x}$ represents the coordinate operator for the single electron which we study.
The forward amplitude is, on the other hand, given by
\begin{eqnarray}
\label{eq:forward}
\langle \Phi_{E_{ec}^I}|e^{-i\sum{\mathbf q_e}\cdot \hat \xe_i}| \Phi_{E_{ec}^I} \rangle
\simeq 1 - \frac{1}{2}\sum_{i=1}^{N_e}
\langle \phi_{o_i} | ( {\mathbf q_e}\cdot\hat\xe )^2
|\phi_{o_i } \rangle 
+\frac{1}{2} \sum_{i,j=1\, (i\neq j)}^{N_e}
\left|\langle \phi_{o_i} |  {\mathbf q_e}\cdot\hat\xe |\phi_{o_j } \rangle\right|^2\, ,
\end{eqnarray}
which leads to the probability with the electron cloud unchanged,
\begin{eqnarray}
\prob_{\rm unchanged}\simeq 
1 - \sum_{i=1}^{N_e}
\langle \phi_{o_i} | ( {\mathbf q_e}\cdot\hat\xe )^2
|\phi_{o_i } \rangle + \sum_{k=1}^{N_e}\sum_{o \in ec_I\backslash \{o_k\}}
\left|\langle \phi_{o} |  {\mathbf q_e}\cdot\hat{\xe}  |\phi_{o_k } \rangle\right|^2\ .
\end{eqnarray}

Let us define probabilities of the single electron transitions,
\begin{eqnarray}
\label{eq:unchanged}
p_{\rm unchanged}(k) &=&\left |\langle \phi_{o_k} | e^{i{\mathbf q_e}\cdot\hat{\xe} }|\phi_{o_k } \rangle \right|^2
\simeq  \left|1- \frac{1}{2}\langle \phi_{o_k} | ( {\mathbf q_e}\cdot\hat\xe )^2|\phi_{o_k } \rangle \right|^2\ , \\
\label{eq:ex}
p_{\rm ex}(k) &=&
\,\,\,\mathclap{\displaystyle\int}\mathclap{\textstyle\sum}\,\,\,_{o\cancel{\in}ec_I}
\left|\langle \phi_{o} | e^{i{\mathbf q_e}\cdot\hat{\xe}} |\phi_{o_k } \rangle \right|^2
\simeq 
\,\,\,\mathclap{\displaystyle\int}\mathclap{\textstyle\sum}\,\,\,_{o\cancel{\in}ec_I}
\left|\langle \phi_{o} | {\mathbf q_e}\cdot\hat{\xe} |\phi_{o_k } \rangle \right|^2\ ,
\\
\label{eq:occupied}
p_{\rm occupied}(k) &=&
\sum_{o{\in} ec_I\backslash\{o_k\} }\left|\langle \phi_{o} | e^{i{\mathbf q_e}\cdot\hat{\xe}} |\phi_{o_k } \rangle \right|^2
\simeq 
\sum_{o{\in} ec_I\backslash\{o_k\} }
\left|\langle \phi_{o} | {\mathbf q_e}\cdot\hat{\xe} |\phi_{o_k } \rangle \right|^2\ ,
\end{eqnarray}
up to ${\cal O}(q_e^3)$.
The probability $p_{\rm occupied}(k)$ denotes the transition between the occupied orbitals.
The single electron transition probabilities satisfy
\begin{eqnarray}
\label{eq:singleUNITARITY}
p_{\rm unchanged}(k) + p_{\rm ex}(k) + p_{\rm occupied}(k) \simeq 
\sum_{o}\left|\langle \phi_{o} | e^{i{\mathbf q_e}\cdot\hat{\mathbf x}} |\phi_{o_k }\rangle \right|^2 = 1 \ .
\end{eqnarray}

To the order of ${\cal O}(q_e^2)$, the total probabilities can be expressed by using the single electron transition probabilities,
\begin{eqnarray}
\prob_{\rm unchanged} + \prob_{\rm ex} &\simeq & \prod_i\left(p_{\rm unchanged}(k) + p_{\rm ex}(k) + p_{\rm occupied}(k)\right)\ .
\end{eqnarray}
Thus, we find that Eq.\,(\ref{eq:singleUNITARITY}) guarantees the probability conservation in terms of the single electron 
transition probabilities.
It should be emphasized that $p_{\rm occupied}$ plays an important role for the conservation of the probability.
As is clear from the above argument, however, $p_{\rm occupied}$ is a part of $\prob_{\rm unchanged}$, and hence, 
it does not contribute to $\prob_{\rm ex}$ up to ${\cal O}(q_e^3)$.

\section{Dirac-Hartree-Fock Method}
\label{sec:DHF}
In this appendix, we briefly review the Dirac-Hartree-Fock method in the natural units
(see~\cite{9783540680109,Gu2008} for review).
The Hamiltonian for the electrons is given by
\begin{equation}
 \hat H_{ec}=\sum_j\hat h_j+\sum_{i<j}\frac{\alpha}{|\hat{\mathbf{r}}_i-\hat{\mathbf{r}}_j|}\, ,
\end{equation}
where
\begin{equation}
 \hat h_j=\bar{\boldsymbol\alpha}\cdot\mathbf p_j+m_e(\beta-1)+V_N(|\hat{\mathbf{r}}_j|)\, .
\end{equation}
Here, we ignore the Breit interaction terms and
\begin{equation}
 V_N(r)=
\begin{cases}
 -\frac{Z\alpha}{2R_N}\left[3-\left(\frac{r}{R_N}\right)^2\right],&r\leq R_N\\
 -\frac{Z\alpha}{r},&r>R_N
\end{cases}\, ,
\end{equation}
with $\alpha m_eR_N=2.2677\times10^{-5}A^{1/3}$. 
The Dirac matrices $\bar{\boldsymbol\alpha}$ and $\beta$ are defined by
\begin{equation}
 \bar{\boldsymbol\alpha}=
\left(\begin{array}{cc}
 0&\boldsymbol\sigma\\
 \boldsymbol\sigma&0 
\end{array}\right)\ , 
\quad \beta=
\begin{pmatrix}
  1_2&0\\
 0&- 1_2 
\end{pmatrix},
\end{equation}
where $\boldsymbol\sigma=(\sigma_1,\sigma_2,\sigma_3)$ are the Pauli matrices.

We assume that the electron wave function for the ground state is approximately given by
\begin{equation}
 \Phi(\mathbf r_1,\cdots,\mathbf r_N)=\sum_{\sigma\in S_{N}}\frac{\mathrm{sgn}(\sigma)}{\sqrt{N!}}\prod_{j=1}^N\phi_{o_{\sigma(j)}}(\mathbf r_j)\, ,
\end{equation}
with
\begin{equation}
 \phi_{o=\{E_{n\kappa},\kappa,m\}}(\mathbf r)=\frac{1}{r}
\begin{pmatrix}
 P_{n\kappa}(r)\Omega_{\kappa m}(\theta,\varphi)\\
 iQ_{n\kappa}(r)\Omega_{-\kappa m}(\theta,\varphi)
\end{pmatrix}\, .
\end{equation}
Here we use $n$ and $\kappa$, instead of $E$, to label the radial wave functions. We choose $\{o_i\}$ to be one of the ground state configurations.%
\footnote{In {\tt FAC}, a fictitious mean configuration with fractional occupation numbers is used if there are more than one ground state configurations.}

The ground state wave functions are determined by the variational method.
The expectation value of the Hamiltonian is calculated as
\begin{align}
 \langle \Phi|\hat H_{ec}|\Phi\rangle&=\sum_j\langle\phi_{o_j}|\hat h|\phi_{o_j}\rangle+\sum_{i<j}\langle\phi_{o_i},\phi_{o_j}|\frac{\alpha}{|\hat{\mathbf{r}}_0-\hat{\mathbf{r}}'_0|}|\phi_{o_i},\phi_{o_j}\rangle\nonumber\\
&\hspace{3ex}-\sum_{i<j}\langle\phi_{o_j},\phi_{o_i}|\frac{\alpha}{|\hat{\mathbf{r}}_0-\hat{\mathbf{r}}'_0|}|\phi_{o_i},\phi_{o_j}\rangle\, ,
\end{align}
where $h$ is the same as $h_j$ but operates on a single electron state.
Here, the expectation values in the coordinate  representation are give by
\begin{align}
 \langle\Phi'|\Phi\rangle&\equiv\sum_{\sigma',\sigma}\frac{\mathrm{sgn}(\sigma')\mathrm{sgn}(\sigma)}{N!}\prod_{j=1}^N\langle\phi_{o'_{\sigma'(j)})}|\phi_{o_{\sigma(j)}}\rangle\, ,\\
 \langle\phi_{o'}|f(\hat{\mathbf{r}}_0)|\phi_{o}\rangle&\equiv\sum_\alpha\int d^3\mathbf r_0[\phi^{\alpha}_{o'}(\mathbf r_0)]^*\phi^\alpha_{o}(\mathbf r_0)f(\mathbf r_0)\, ,\\
 \langle\phi_{o_1},\phi_{o_2}|f(\hat{\mathbf{r}}_0,\hat{\mathbf{r}}'_0)|\phi_{o_3},\phi_{o_4}\rangle&\equiv\sum_{\alpha,\beta}\int d^3\mathbf r_0d^3\mathbf r'_0[\phi^{\alpha}_{o_1}(\mathbf r)\phi^{\beta}_{o_2}(\mathbf r'_0)]^*\phi^\alpha_{o_3}(\mathbf r_0)\phi^\beta_{o_4}(\mathbf r'_0)f(\mathbf r_0,\mathbf r'_0)\, .
\end{align}

By taking the variation for the coordinate of one of the orbital electrons,
we obtain
\begin{align}
 0&=\hat h\phi_{o_j}(\mathbf r)+\left[\sum_{i(i\neq j)}\langle\phi_{o_i}|\frac{\alpha}{|\mathbf r-\hat{\mathbf{r}}_0|}|\phi_{o_i}\rangle\right]\phi_{o_j}(\mathbf r)\nonumber\\
&\hspace{3ex}-\left[\sum_{i(i\neq j)}\phi_{o_i}(\mathbf r)\langle\phi_{o_i}|\frac{\alpha}{|\mathbf r-\hat{\mathbf{r}}_0|}\right]|\phi_{o_j}\rangle-\varepsilon_{o_j}\phi_{o_j}(\mathbf r)+h.c.\, ,\label{eq:DHF}
\end{align}
where $\varepsilon_o$ is a Lagrange multiplier to impose 
\begin{equation}
 \langle\Phi|\Phi\rangle=1\, .
\end{equation}
This is the so-called Dirac-Hartree-Fock equation and gives simultaneous differential equations for $\phi_{o_j}$. 
The second and the third terms express the electron-electron interaction and can be seen 
as local and non-local potentials for $\phi_{o_j}$, respectively, once $\phi_{o_i}$ are treated as mean fields.

Since the non-local potential is numerically demanding, the Slater approximation is often adopted to localize the potential~\cite{Kohn:1965zzb,Sampson1989}. 
However, since it has incorrect asymptotic behavior, {\tt FAC} uses an improved potential given in Eq.\,(\ref{eq:FACpotential}).
From Eq.\,\eqref{eq:DHF}, we finally obtain,
\begin{align}
 \left(\frac{d}{dr}+\frac{\kappa}{r}\right)P_{n\kappa}(r)&=\left(\varepsilon_{n\kappa}-V(r)+2m_e\right)Q_{n\kappa}(r)\, ,\label{eq:DHF1}\\
 \left(\frac{d}{dr}-\frac{\kappa}{r}\right)Q_{n\kappa}(r)&=\left(-\varepsilon_{n\kappa}+V(r)\right)P_{n\kappa}(r)\, ,\label{eq:DHF2}
\end{align}
with
\begin{equation}
 V(r)=V_N(r)+V_{ee}(r)\, ,
\end{equation}
which is solved iteratively for the ground state.
As for the excited states and the unbouded states, the single electron wave functions are obtained by solving Eqs.\,\eqref{eq:DHF1} and \eqref{eq:DHF2} by using the potential in Eq.\,(\ref{eq:FACpotential}), which is iteratively obtained for the ground state.

\bibliography{draft_arxiv_revise2}
\end{document}